\documentclass[sigconf,authorversion,nonacm]{acmart}
\AtBeginDocument{%
  }

\copyrightyear{2026}
\acmYear{2026}
\acmConference[CHI '26]{Proceedings of the 2026 CHI Conference on Human Factors in Computing Systems}{April 13--17, 2026}{Barcelona, Spain}
\acmBooktitle{Proceedings of the 2026 CHI Conference on Human Factors in Computing Systems (CHI '26), April 13--17, 2026, Barcelona, Spain}
\acmDOI{10.1145/3772318.3790974}
\acmISBN{979-8-4007-2278-3/2026/04}

\acmSubmissionID{4136}

\usepackage{algorithm}
\usepackage{makecell}
\usepackage{multirow}

\providecommand{\mytoolname}{{\textit{Cerebra}}}
\providecommand{\usrfb}[1]{{``\textit{#1}''}}

\definecolor{codebgcolor}{gray}{0.95} 
\definecolor{picolor}{RGB}{36,175,255}
\definecolor{krcolor}{RGB}{0,106,183}
\definecolor{socolor}{RGB}{255,144,116}
\definecolor{kicolor}{RGB}{203,31,3}

\newcommand{\ilogpi}[1]{\textbf{\textcolor{picolor}{#1}}}
\newcommand{\ilogkr}[1]{\textbf{\textcolor{krcolor}{#1}}}
\newcommand{\ilogso}[1]{\textbf{\textcolor{socolor}{#1}}}
\newcommand{\ilogki}[1]{\textbf{\textcolor{kicolor}{#1}}}

\providecommand{\DIFadd}[1]{{\protect\color{black}#1}}
\providecommand{\DIFdel}[1]{}

\providecommand{\CAMadd}[1]{{\protect\color{black}#1}}
\providecommand{\CAMdel}[1]{}

\begin{document}

\title{Cerebra: Aligning Implicit Knowledge in Interactive SQL Authoring}

\author{Yunfan Zhou}
\orcid{0009-0009-7814-4390}
\affiliation{%
    \institution{State Key Lab of CAD\&CG\\Zhejiang University}
    \city{Hangzhou}
    \state{Zhejiang}
    \country{China}
}
\email{yf.zhou@zju.edu.cn}

\author{Qiming Shi}
\orcid{0009-0002-7876-1056}
\affiliation{%
    \institution{State Key Lab of CAD\&CG\\Zhejiang University}
    \city{Hangzhou}
    \state{Zhejiang}
    \country{China}
}
\email{qimingshi@zju.edu.cn}

\author{Zhongsu Luo}
\orcid{0009-0003-0885-2742}
\affiliation{%
    \institution{State Key Lab of CAD\&CG\\Zhejiang University}
    \city{Hangzhou}
    \state{Zhejiang}
    \country{China}
}
\email{zhongsuluo@zju.edu.cn}

\author{Xiwen Cai}
\orcid{0000-0002-7256-4660}
\affiliation{%
    \institution{Department of Digital Intelligence\\China Mobile}
    \city{Shenzhen}
    \state{Guangdong}
    \country{China}
}
\email{caixiwen@chinamobile.com}

\author{Yanwei Huang}
\orcid{0009-0001-9453-7815}
\affiliation{%
    \institution{HKUST}
    \city{Hong Kong S.A.R.}
    \country{China}
}
\email{yanwei.huang@connect.ust.hk}

\author{Dae Hyun Kim}
\orcid{0000-0002-8657-9986}
\affiliation{%
    \institution{Department of Computer Science and Engineering, Yonsei University}
    \city{Seoul}
    \country{Republic of Korea}
}
\affiliation{%
    \institution{Graduate School of Artificial Intelligence, POSTECH}
    \city{Pohang}
    \country{Republic of Korea}
}
\email{dhkim16@yonsei.ac.kr}

\author{Di Weng}
\orcid{0000-0003-2712-7274}
\authornote{Di Weng is the corresponding author.}
\affiliation{%
    \institution{School of Software Technology\\Zhejiang University}
    \city{Ningbo}
    \state{Zhejiang}
    \country{China}
}
\email{dweng@zju.edu.cn}

\author{Yingcai Wu}
\orcid{0000-0002-1119-3237}
\affiliation{%
    \institution{State Key Lab of CAD\&CG\\Zhejiang University}
    \city{Hangzhou}
    \state{Zhejiang}
    \country{China}
}
\email{ycwu@zju.edu.cn}

\renewcommand{\shortauthors}{Zhou et al.}

\begin{abstract}
  LLM-driven tools have significantly lowered barriers to writing SQL queries.
However, user instructions are often underspecified, assuming the model understands implicit knowledge, such as dataset schemas, domain conventions, and task-specific requirements, that isn't explicitly provided.
This results in frequently erroneous scripts that require users to repeatedly clarify their intent.
Additionally, users struggle to validate generated scripts because they cannot verify whether the model correctly applied implicit knowledge.
We present \textit{Cerebra}, an interactive NL-to-SQL tool that aligns implicit knowledge between users and LLMs during SQL authoring.
\textit{Cerebra} automatically retrieves implicit knowledge from historical SQL scripts based on user instructions, presents this knowledge in an interactive tree view for code review, and supports iterative refinement to improve generated scripts.
To evaluate the effectiveness and usability of \textit{Cerebra}, we conducted a user study with 16 participants, demonstrating its improved support for customized SQL authoring.
The source code of \textit{Cerebra} is available at \url{https://github.com/zjuidg/CHI26-Cerebra}.
\end{abstract}



\begin{teaserfigure}
  \includegraphics[width=\textwidth]{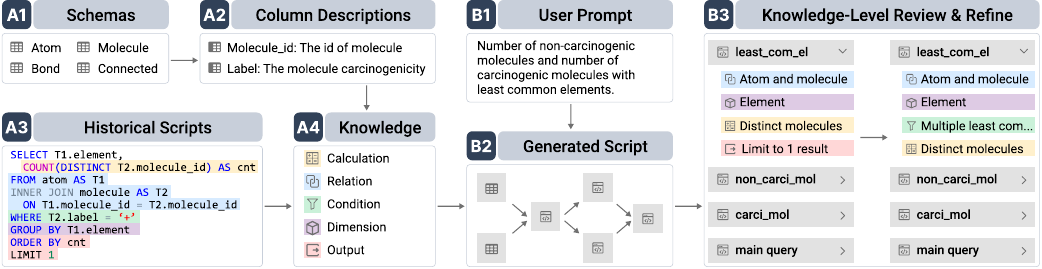}
  \caption{The workflow of \mytoolname\ consists of two stages: offline (A1-A4) and online (B1-B3). In the offline stage, \mytoolname\ first generates column descriptions (A2) based on the database schemas (A1). It then extracts five types of implicit knowledge (A4) in natural language from user-authored historical scripts (A3). In the online stage, when a user submits a natural language instruction (B1), \mytoolname\ retrieves relevant knowledge items and generates the corresponding SQL script (B2). The generated script is then parsed and presented in the Knowledge View (B3), where users can review and refine the knowledge items to iteratively improve the script.}
  \Description{Cerebra workflow diagram with two stages: an offline stage (A1-A4) and an online stage (B1-B3). In the offline stage, A1 shows database schemas with tables for Atom, Molecule, Bond, and Connected; A2 displays column descriptions such as ``Molecule\_id: The id of molecule'' and ``Label: The molecule carcinogenicity''; A3 presents a sample SQL query in color-coded syntax representing a historical user script; A4 lists five types of extracted knowledge---calculation, relation, condition, dimension, and output---each in a colored box. In the online stage, B1 is a user prompt asking for counts of non-carcinogenic and carcinogenic molecules with the least common elements; B2 illustrates the generation of a SQL script using connected nodes and links; B3 shows a knowledge-level review interface, where knowledge items like ``atom and molecule'', ``element'', ``distinct molecules'', ``limit to 1 result'', and ``multiple least common element'', are displayed and can be refined, with updates reflected in the organization of knowledge items for different subqueries.}
  \label{fig:cerebra-workflow}
\end{teaserfigure}

\maketitle

\section{Introduction}
\label{sec:intro}
With the rapid growth of available data in various fields, being able to explore data effectively with SQL queries has become crucial for data practitioners.
\DIFdel{While those who have fluent programming skills can write SQL scripts to aid their exploration, not all data practitioners are able to write code to clearly express their intents. }
\DIFadd{Even for those who are proficient in SQL, authoring code that clearly expresses their intents can be difficult and time‑consuming~\cite{queryvis}, while many other practitioners have only limited programming experience~\cite{sqlucid}.}
As a result, Natural Language to SQL (NL-to-SQL) techniques, especially LLM-driven ones~\cite{codes, din-sql, chat2db} have been widely adopted to lower the barriers to query construction~\cite{nl2sql360}.
Recent advances in NL-to-SQL techniques have led to substantial improvements in basic semantic understanding and translation~\cite{nl2sql-survey, nl2sql-survey-dl}, 
with state-of-the-art models achieving over 90\% accuracy on the Spider benchmark~\cite{spider} for well-formed natural language instructions and database schemas.

\begin{figure*}[ht]
  \centering
  \includegraphics[width=\linewidth]{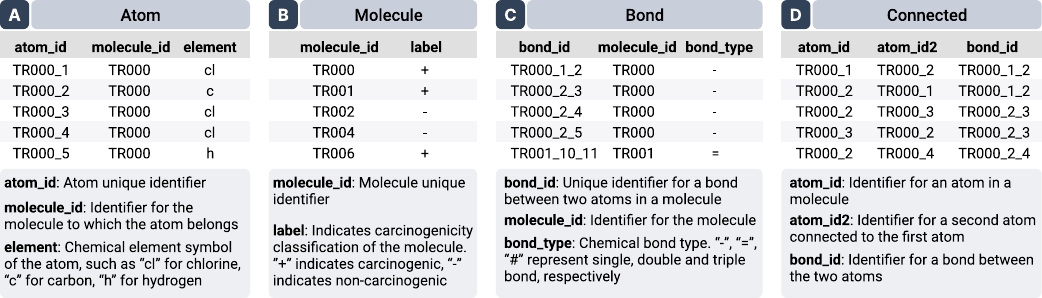}
  \caption{The schema, sample data and column descriptions for the Toxicology database in a NL-to-SQL dataset, BIRD. The database consists of four tables: A) Atom, which records the atoms present in each molecule. B) Molecule, which stores the properties of each molecule, including its carcinogenicity label. C) Bond, which describes the chemical bonds between atoms within a molecule. D) Connected, which represents the connectivity relationships between pairs of atoms via specific bonds.}
  \Description{Schemas of four tables from the Toxicology database in the BIRD NL2SQL dataset: Atom, Molecule, Bond, and Connected. The Atom table lists atom\_id, molecule\_id, and element, with sample data such as atom TR000\_1 in molecule TR000 as chlorine (``cl''); Molecule contains molecule\_id and label, where the label ``+'' indicates carcinogenic and ``-'' indicates non-carcinogenic; Bond includes bond\_id, molecule\_id, and bond\_type, with bond\_type symbols ``-'', ``='', and ``\#'' for single, double, and triple bonds; Connected lists atom\_id, atom\_id2, and bond\_id for pairs of atoms connected by specific bonds. Each table is accompanied by brief column descriptions explaining the meaning of each field.}
  \label{fig:toxicology-schema}
\end{figure*}

Despite recent advances, users continue to struggle with obtaining accurate results from NL-to-SQL systems when their natural language queries \DIFdel{rely on}\DIFadd{involve} \textit{implicit knowledge}.
\DIFadd{This refers to the assumptions, conventions, and contexts that users do not articulate explicitly but are crucial to accurate query generation.}
Such knowledge encompasses dataset-specific conventions and task-specific computations\DIFdel{that users do not explicitly articulate in their instructions}, creating a knowledge alignment gap that hinders accurate query interpretation and SQL generation.
The BIRD toxicological database~\cite{bird} (\autoref{fig:toxicology-schema}) exemplifies this issue.
A query like ``\textit{List all non-bonding elements}'' assumes implicit knowledge that non-bonding elements are atoms in the ``Atom'' table with no corresponding entries in the ``Connected'' table.
Typical NL-to-SQL models, lacking this understanding, may fail to generate the correct SQL query.
Another example involves queries like ``\textit{What is the difference between the number of carcinogenic molecules and the number of non-carcinogenic molecules}''.
Such queries also assume implicit dataset-specific knowledge that carcinogenicity status is represented by labels such as ``+'' and ``-'' in a specific field.
As a result, users frequently need to repeatedly clarify implicit knowledge such as the meaning of ``non-bonding elements'' and ``carcinogenic'' status, leading to frustration from iterative query refinement and numerous unexpected failures.

To address this gap, we conducted a preliminary study to investigate the role of implicit knowledge in the SQL authoring process. Through interviews with 10 data practitioners who regularly use LLM-driven NL-to-SQL tools, we found that users consistently supplement their natural language queries with extensive contextual information, including database schemas and business backgrounds, to bridge implicit knowledge gaps. Many participants (6/10) also reused existing SQL scripts in their prompts, effectively transferring the implicit domain logic and computational patterns embedded within proven solutions.
However, users face a fundamental tension: concise queries fail because they omit essential implicit assumptions, while making all implicit knowledge explicit through detailed prompts becomes excessively burdensome and error-prone. Furthermore, users struggle to interpret the implicit assumptions that LLMs make when generating code, finding it difficult to align high-level requirements with low-level code fragments and locate errors when their implicit expectations are not met.
While prior research has explored interactive interfaces for SQL construction, understanding, and modification~\cite{nl-edit, datatone, eviza, sqlucid}, the challenge of surfacing and aligning implicit knowledge in users' natural language instructions remains largely unaddressed. 
These findings highlight the need for NL-to-SQL tools that can better surface and align implicit knowledge between users and models.

In response to these challenges, we present \mytoolname,
an interactive NL-to-SQL tool designed to surface and align implicit knowledge between users and LLMs during SQL authoring.
Drawing directly from our preliminary study findings, particularly participants' practice of reusing existing scripts to transfer implicit knowledge, \mytoolname\ systematically captures and leverages the implicit knowledge embedded in users' historical SQL scripts.
We first introduce a framework that categorizes implicit knowledge in NL-to-SQL processes into five types: calculation, condition, relation, dimension, and output.
\mytoolname\ extracts natural language descriptions and associated code fragments from users' past queries based on these categories, creating a repository of implicit knowledge patterns.
During query authoring, \mytoolname\ retrieves relevant implicit knowledge from this repository to enhance SQL generation, addressing the core challenge of conveying unstated assumptions from users to LLMs.
To tackle the interpretation challenge identified in our study, \mytoolname\ features a knowledge tree view that visualizes how the system inferred and applied implicit knowledge in generating SQL scripts.
This transparency allows users to understand the model's assumptions and iteratively refine both the implicit knowledge and the generated SQL when results don't meet their expectations.

To evaluate the usability and effectiveness of \mytoolname, we conducted a counterbalanced mixed-design user study on 16 participants.
Participants of the user study completed the SQL authoring tasks within a significantly shorter period of time  with \mytoolname\ than a baseline tool without implicit knowledge integration.
User feedback further confirms that \mytoolname\ helps users convey their implicit knowledge in natural language more easily and better understand the implicit knowledge inferred by the model.
The key contributions of this paper are three-fold:

\begin{itemize}
  \item A preliminary study with 10 data practitioners that reveals the central role of implicit knowledge in NL-to-SQL tasks, identifying key challenges in implicit knowledge transfer between humans and LLMs.
  \item \DIFdel{\mbox{\mytoolname}, an interactive NL-to-SQL tool that leverages historical scripts to capture implicit knowledge and provides transparent knowledge visualization to improve human-LLM alignment in SQL authoring.}\DIFadd{\mytoolname, an interactive NL-to-SQL tool that captures implicit knowledge from historical scripts, decomposes user requests into subqueries, and provides transparent knowledge visualization to improve human-LLM alignment in SQL authoring.}
  \item A comparative evaluation demonstrating the usability and effectiveness of \mytoolname\ in supporting communication of implicit knowledge in NL-to-SQL tasks.
\end{itemize}

\section{Related Work}
\label{sec:rw}
In this section, we review previous research on interactive SQL authoring, NL-driven data querying, and human-AI collaboration.

\subsection{Interactive SQL Authoring Tools}
\label{ssec:rw-sql-tools}

Authoring SQL queries is a common practice in database development.
Traditional integrated development environments (IDEs) like Navicat~\cite{navicat} and DataGrip~\cite{datagrip} provide basic SQL authoring functionalities such as code completion and code formatting.
Since these IDEs require significant expertise to use effectively, several \textit{visual programming} tools~\DIFdel{\mbox{\cite{sqlsnap, tioga-2}}}\DIFadd{\cite{sqlsnap, tioga-2, dataplay, asyncflow, aflak}} are designed to lower the barrier \DIFadd{by providing tree or graph representation of the code}.
For instance, Tioga-2~\cite{tioga-2} incorporates a data flow diagram to represent SQL queries and users can directly manipulate nodes and links to edit the queries.
However, these visual programming tools often rely on predefined templates, making it demanding to configure numerous options to construct queries.

To better support SQL authoring requirements, many SQL authoring tools integrate program synthesis techniques like \textit{programming by example}~\cite{scythe, sqlsynthesizer, qfe} or \textit{NL-to-SQL}~\cite{din-sql, codes, dail-sql}, where users can declaratively specify their query requirements.
Specifically, there is a substantial amount of research focusing on NL-to-SQL models, adopting rule-based~\cite{nalir, athena, sqlizer}, neural network-based~\cite{irnet, rat-sql, syntaxsqlnet}, pre-trained language model-based~\cite{resdsql, graphix-t5, picard} and LLM-based~\cite{din-sql, codes, dail-sql} methods.
See~\cite{nl2sql-survey} for a comprehensive survey of these models.
The advancements in NL-to-SQL techniques enable SQL authoring tools to understand more complex natural language and table structures in sophisticated databases, thereby providing users with smarter code suggestions.
However, these NL-to-SQL models cannot always provide accurate results that satisfy user needs, requiring considerable effort in validating the generated queries and correcting the mistakes.
To bridge the gap, numerous interactive tools are proposed,\DIFdel{to support \textit{understanding} and \textit{refining} the generated SQL queries.}
\DIFadd{many of which do not integrate LLMs, but instead focus on facilitating two key developer tasks: \textit{code understanding} and \textit{query refinement}.}

\DIFdel{There is a line of work that focuses on code \textit{understanding}
by visualizing the abstract syntax trees (ASTs)~\mbox{\cite{sqlvis, queryvis, queryscope}}, providing intermediate results~\mbox{\cite{data-tweening, diy, sqlucid, i-rex}}, or offering natural language explanations~\mbox{\cite{logos, diy, sqlucid, explain-sql-nl}}.
For instance, QueryVis~\mbox{\cite{queryvis}} translates SQL queries into diagrams based on first-order logic principles, exposing the underlying logical patterns of queries.
Data tweening~\mbox{\cite{data-tweening}} incrementally interpolates between SQL result sets to create smooth transitions during interactive query sessions, helping users better understand complex data transformations.
DIY~\mbox{\cite{diy}} provides users with a sandbox environment to interact with mappings between natural language queries and SQL queries, assisting end users in validating generated scripts.}
\DIFadd{\textit{Code understanding} tools aim to demystify the logic of SQL queries by employing four primary strategies.
First, to abstract away complex syntax, several tools focus on visualizing logical structure through delicately designed graphs~\cite{sqlvis, queryvis, queryscope, polaris, sflvis}.
For instance, QueryVis~\cite{queryvis} translates SQL queries into diagrams based on first-order logic principles to reveal underlying logical patterns.
Second, several tools facilitate inspecting intermediate execution states~\cite{data-tweening, i-rex, habitat, xavier, codelin}.
For example, Data Tweening~\cite{data-tweening} visualizes incremental transformations between result sets.
Third, to facilitate easy code validation,
tools such as DIY~\cite{diy} and RATest~\cite{ratest} generate small sample databases or counterexamples in a sandbox environment.
Fourth, several approaches utilize natural language generation to translate complex query logic into readable narratives~\cite{logos, explain-sql-nl, sqlucid, steps}.
Systems like STEPS~\cite{steps} and SQLucid~\cite{sqlucid} decompose SQL queries into clauses and provide step-by-step explanations.}
In addition, extensive research has been conducted in code debugging and we refer readers to the relevant surveys~\cite{debug-sql-survey, software-vis-survey}.
Whereas the aforementioned tools provide various methods to assist users in syntactically understanding the logical structure and the execution output of SQL queries,
our work emphasizes semantic understanding by presenting the implicit knowledge underlying the generation process.
This allows users to comprehend the assumptions and conventions reflected in the generated queries.

Another line of work provides interactive support for \DIFdel{\textit{refining} the generated queries}\DIFadd{\textit{query refinement}}, which can be categorized into \textit{NL-based} and \textit{GUI-based} approaches.
In \textit{NL-based} tools, users can edit the queries by answering multi-choice clarification questions~\cite{dialsql, piia, misp}, modifying step-by-step NL explanations~\cite{sqlucid}, or using free-form text~\cite{nl-edit}.
For example, DialSQL~\cite{dialsql} leverages a hierarchical encoder-decoder architecture to predict error categories, locate error spans in the query, and generate candidate choices for users to select from.
SQLucid~\cite{sqlucid} proposes a fine-grained query refinement approach where users can modify steps in the NL explanation to correct errors at the clause or entity level.
NL-EDIT~\cite{nl-edit} interprets natural language corrections provided by users to generate a sequence of edits that can be applied to the initial SQL query.
\textit{GUI-based} tools enable users to make simple query edits by providing widgets like menus~\cite{datatone, nalir} and sliders~\cite{eviza, orko}.
For instance, in DataTone~\cite{datatone}, users can interact with the dynamically generated ambiguity widgets to correct table names, column names and data values in the query.
\DIFadd{While these refinement techniques help resolve ambiguity in natural language specifications, they generally do not capture or reuse implicit knowledge about dataset-specific conventions and task-specific computations, which our work explicitly models and exposes at the knowledge level.}

\DIFdel{The studies mentioned above have introduced diverse interactive support to address the ambiguity issue of natural language while generating queries, democratizing SQL authoring for users.
However, existing tools often lack mechanisms for capturing and utilizing users' implicit knowledge.
Users are frequently required to repeatedly explain their dataset-specific conventions and task-specific computations in their prompts.
In contrast, our work enables the automatic capture and reuse of implicit knowledge from historical scripts, and further supports knowledge-level refinement, allowing users to directly adjust and validate the knowledge that guides SQL generation.}

\DIFadd{Beyond SQL-specific tools, we draw inspiration from recent work on interactive NL-driven code authoring, which explores decomposing tasks into editable intermediate artifacts~\cite{stepwise-phasewise, neurosync, coladder}.
Systems like Phasewise and Stepwise~\cite{stepwise-phasewise} decompose AI-generated solutions into linear \textit{lists} of editable assumptions and executable plans to support stepwise steering.
CoLadder~\cite{coladder} and NeuroSync~\cite{neurosync} expose structured and editable intent representations through hierarchical prompt \textit{trees} and intent \textit{graphs}, enabling programmers to externalize their goals and to refine inferred tasks.
Inspired by these methods, \mytoolname\ adopts a hybrid approach to query decomposition,
combining a \textit{graph}-based data flow diagram that reveals subquery dependencies with a \textit{tree}-structured Knowledge View that provides code-knowledge highlighting for precise inspection and modification of the generated query.}

\subsection{Interactive NL-Driven Data Querying Tools}

\DIFadd{Beyond SQL generation, interactive data querying tools employ natural language interfaces (NLIs) to facilitate visual data exploration, addressing four primary interaction barriers when using natural language.
To address the \textit{abstraction} barrier inherent in describing high-level descriptive concepts (e.g., shapes, trends, patterns),
tools such as SlopeSeeker~\cite{slopeseeker} and ShapeSearch~\cite{shapesearch} allow users to query quantifiable trends and complex shapes by mapping qualitative adjectives to mathematical definitions.
To overcome the \textit{imprecision} barrier caused by vague terms (e.g., ``large'', ``many'', or ``near''),
systems like DataTone~\cite{datatone} and Eviza~\cite{eviza} utilize ambiguity widgets to help users select their intended data attributes and values,
while approaches like \cite{nl-query-traj} leverage multiple views to enable fuzzy spatial constraints for querying uncertain trajectory data.
To mitigate the \textit{formulation} barrier where users know what they exactly want but do not know how to use the system to form valid queries,
Sneak Pique~\cite{sneak-pique} provides data-aware autocompletion to preview results,
NL2Rigel~\cite{nl2rigel} maps natural language instructions to data transformation specifications,
and NL4DV Toolkit~\cite{nl4dv} abstracts natural language processing complexities to aid developers in creating data querying interfaces.
To lower the \textit{continuity} barrier in maintaining conversational context,
Evizeon~\cite{evizeon} incorporates pragmatics to handle pronouns and incomplete utterances, whereas Orko~\cite{orko} manages context across multimodal inputs like touch and speech.}

\DIFadd{While the aforementioned NLIs primarily address ambiguity by clarifying the operations and data references, they often overlook the \textit{knowledge alignment} barrier, which is the gap between the user's domain-specific assumptions and the model's understanding.
Unlike abstraction or imprecision barriers which arise from certain terms in the user instruction, this new barrier involves the implicit dataset-specific conventions and task-specific computations that users frequently omit from their instructions.}

\subsection{Human-AI Collaboration}
\label{ssec:human-ai-collaboration}

\DIFadd{\textbf{Empirical Studies of SQL Authoring}.
A growing body of work has examined how people author and debug SQL, both in traditional environments~\cite{sqlshare, sqlvis} and with NL-to-SQL tools~\cite{model-user-err-nldb, nl2sql-industry}.
For instance, SQLShare's deployment logs and interviews show that scientists routinely reuse and layer views over user‑uploaded tables to encode domain logic and processing pipelines~\cite{sqlshare}.
For NL-to-SQL specifically, Ning et al.~\cite{model-user-err-nldb} derive a detailed taxonomy of NL-to-SQL model errors and demonstrate that end users have difficulty both detecting these errors and repairing them reliably through a user study.
Building on these findings, several interactive systems study how explanations and direct manipulation affect SQL authoring~\cite{diy, steps, sqlucid}.
These works characterize the challenges humans face in understanding and correcting SQL and NL‑to‑SQL outputs.
In contrast, our work focuses on surfacing and aligning implicit knowledge that underlies those outputs, rather than merely their syntactic correctness.}

\DIFadd{\textbf{Capturing Implicit Intentions in Human-LLM Interaction}.
Recent work has begun to examine how people surface and refine implicit intentions when interacting with LLMs.
One line of systems~\cite{stepwise-phasewise, who-validates-the-validators, jupybara} emphasizes structured task decomposition in text, helping users iteratively articulate assumptions and plans while seeing how those choices shape model behavior.
A second line~\cite{amplio, visegpt, neurosync} focuses on visual or GUI-based representations of the model’s internal reasoning or search space, so that users can better inspect and adjust what the model “thinks” it is doing.
A third line~\cite{imaginationvellum, sketchgpt, code-shaping} explores multimodal inputs as a more natural substrate for intent expression, allowing people to use sketches, spatial layouts, or other modalities to convey expectations that are hard to verbalize.
Building on these research directions, our work extends them to the NL-to-SQL setting by modeling the implicit knowledge behind SQL queries, and by providing an interface where users can directly inspect and iteratively refine that knowledge.}

\DIFdel{Human-AI collaboration has become a central topic in the design of intelligent systems, especially in complex tasks such as NL-to-SQL authoring.
Early work on man-computer symbiosis~\mbox{\cite{man-computer}} envisioned computers performing routine operations to support human insight, while humans contribute domain expertise for decision-making.
In recent research, three key forms of collaboration have emerged~\mbox{\cite{retrolens}}:
AI-assisted decision-making, where AI provides recommendations and humans retain control~\mbox{\cite{does-the-whole, human-ai-deliberation, amuse}};
human-in-the-loop paradigms, where humans iteratively review, correct, or refine AI outputs to improve system performance~\mbox{\cite{clinical-decision-making, instructpipe}};
and joint action, in which humans and AI coordinate as a team, dynamically allocating tasks to leverage their respective strengths~\mbox{\cite{cooperative-game-setting, ideationweb}}.
Recent studies have shown that effective collaboration requires not only technical accuracy from AI models but also transparency, clear communication, and appropriate task allocation between human and machine agents~\mbox{\cite{stop-explain}}.}

\DIFdel{Despite advances in LLMs and intelligent assistants, challenges remain in aligning AI-generated outputs with users’ implicit knowledge and intentions.
A major barrier is the lack of interpretability.
Users often struggle to determine which requirements are satisfied by AI-generated code, and where errors originate~\mbox{\cite{what-it-wants-me-to-say}}.
Prior work has explored techniques such as interactive explanations, confidence indicators, and context-aware visualizations to help users understand and control AI behavior~\mbox{\cite{effect-of-confidence, eviza, sqlucid}}.
However, in the context of NL-to-SQL, existing tools rarely make explicit the implicit knowledge inferred by the model, making it difficult for users to validate and iteratively refine generated queries.
Our work builds on these insights by proposing \mbox{\mytoolname} that surfaces implicit knowledge and supports knowledge-level refinement, aiming to foster more transparent and effective human-AI collaboration in SQL authoring.}

\section{Preliminary Study}
\label{sec:preliminary}
To effectively develop LLM-driven NL-to-SQL tools, it is essential to understand the unique challenges and requirements faced by users.
Prior research~\cite{diy, sqlucid, model-user-err-nldb} conducted user studies of NL-to-SQL tools and uncovered key requirements involving query understanding, error detection, and interactive repair. 
These findings have highlighted directions for enhancing LLM-driven NL-to-SQL tools.
However, the role of implicit knowledge in SQL authoring has not been thoroughly investigated.
To bridge this gap, we conducted a preliminary study\footnote{The study has received approval from State Key Lab of CAD\&CG, Zhejiang University.} involving participants expert in SQL authoring.
\DIFdel{In a semi-structured interview (Section~\mbox{\ref{ssec:ps-procedure}}), participants shared their experiences composing SQL scripts using LLM-driven NL-to-SQL tools and discussed the issues they encountered.
We identified key findings (Section~\mbox{\ref{ssec:findings}}) and summarized user requirements (Section~\mbox{\ref{ssec:user-requirements}}) from the interview.}

\subsection{Participants}
\label{ssec:ps-participants}

We recruited 10 data practitioners (denoted as P1-P10, 9 male and 1 female, $Age_{mean} = 35.6$, $Age_{std} = 5.40$) by sending invitations via social media.
They were from industry with diverse backgrounds such as Geographic Information System, Enterprise Resource Planning, and Stock Market Analysis.
All participants were expert in SQL programming ($Experience_{mean} = 11.5$ years, $Experience_{std} = 6.02$ years) and regularly authored queries in their projects (at least once a week).
They also reported moderate familiarity with LLM-driven NL-to-SQL tools ($M = 4.7$) on a 7-point Likert scale (1 = not at all familiar, 7 = extremely familiar).
\DIFdel{Their detailed demographic information is left to the supplementary materials.}
\DIFadd{Their detailed demographic information is provided in \autoref{tab:demographics-preliminary} in the appendix.}
Participants consented to having their shared SQL scripts and their voices recorded.

\subsection{Procedure}
\label{ssec:ps-procedure}

Prior to each interview, we provided participants with a brief overview of the study, including its purpose, procedure, and compensation, before obtaining their consent for data collection.
During the one-on-one, semi-structured interviews, we first gathered background information on participants' work domains, typical tasks, and dataset characteristics.
We then asked participants to describe at least one recent experience of using LLM-driven NL-to-SQL tools to author SQL scripts, encouraging them to walk through the process and discuss any difficulties or issues encountered.
Furthermore, we followed up by discussing specific challenges or interesting behaviors that emerged during the interview.
All the interviews were audio-recorded for subsequent analysis.
The entire interview lasted around 40 minutes and each participant received 50 Chinese Yuan as compensation for their time. 

\subsection{Findings}
\label{ssec:findings}

We recorded all interviews and transcribed the audio recordings into text.
To analyze the user feedback, we conducted an inductive content analysis~\cite{content-analysis}.
The first author initially reviewed the recording transcripts to identify the relevant comments.
Then similar comments were grouped into several themes.
Finally, the authors reviewed the transcripts and discussed the themes on weekly meetings to reach the consensus about the key findings.
In this subsection, we summarize the key findings regarding the general workflow and the challenges participants faced from the interview.

\subsubsection{How Do People Work with LLM-Driven NL-to-SQL Tools to Author SQL Scripts?}
\label{sssec:how-work-with-nl2sql}

Drawing from the interview and participants' demonstrations of recent usage examples of LLM-driven NL-to-SQL tools, we observed a three-stage workflow: 
participants begin with \textit{prompt formulation} where they craft initial queries and provide contextual information, followed by \textit{code review} where they validate the generated SQL against their expectations, and conclude with \textit{code refinement} where they iteratively correct and improve the results when discrepancies are found.
The similar workflows are also observed in prior studies~\cite{waitgpt}. 

\textbf{All participants supplemented contextual information in addition to their natural language queries.}
In the \textit{prompt formulation} stage, participants recognized that their natural language requests alone might be insufficient and consistently provided additional context to bridge potential knowledge gaps.
This supplementation took multiple forms: participants explicitly included database schemas (P1-P6, P9, P10) to clarify structural relationships and table connections that their queries assumed but didn't state. 
They also provided business backgrounds (P1, P3-P9) to ensure the model understood domain-specific terminology and conventions within their organizational context. 
For example, P3 noted that he would send a paragraph from a data report he was working on to ChatGPT for business context.
The need for context supplementation highlights how natural language queries alone are insufficient, and participants must anticipate and address implicit knowledge gaps. 

\textbf{Participants preferred to reuse existing scripts to help LLMs learn implicit knowledge embedded within them.}
We noticed that more than half of the participants (6/10) tended to incorporate previously written queries into the prompt to obtain queries that aligned with the provided ones (P1, P4, P6-P9).
P1 pointed out a scenario for reuse,
\usrfb{I am writing queries periodically within a specific domain, my query logic is fairly fixed -- only minor adjustments to parameters or subqueries}.
Other participants highlighted the benefits of reuse.
First, existing scripts contain knowledge about complex data transformations, filtering conditions, and computational formulas that would require extensive explanation if articulated from scratch.
As P4 described, \usrfb{I usually find a script I wrote before and tweak it using AI -- it’s much faster that way.}
Second, since previously written scripts embodied well-tested domain logic, reusing scripts inherited this implicit knowledge while reducing validation overhead, which echoed prior research~\cite{strategies-reuse}.
As P6 noted, \usrfb{I don't feel confident using AI to write a SQL script from scratch, so I typically reuse existing queries in the prompt.}

\subsubsection{What Makes LLM-Driven NL-to-SQL Tools Challenging to Use in Practice?}
We organize findings around two primary challenges of implicit knowledge transfer between humans and LLMs.

\textbf{Challenges in conveying implicit knowledge (human $\rightarrow$ LLM).}
In the \textit{prompt formulation} stage, participants (7/10) reported that when they omitted details they viewed as \DIFdel{``common sense''}\DIFadd{conventions or practices} within their domain, LLMs typically failed to accurately interpret these instructions.
\DIFdel{This ``common sense''}\DIFadd{Such details} can be categorized into data-related and computation-related knowledge.
The data-related knowledge mentioned by participants involved data values meanings
(e.g., \usrfb{A value of 1 denotes offline sales, while a value of 2 indicates online sales.} (P3)),
data format constraints (e.g., \usrfb{The string length of this attribute is no more than 20.} (P8)),
and table relations (e.g., \usrfb{These eight tables are intricately interconnected.} (P2)).
The computation-related knowledge involved alias-naming standards 
(e.g., \usrfb{I won't give aliases of derived attributes in that way.} (P7)),
filtering conditions (e.g., \usrfb{I would avoid fuzzy matches to filter data -- they produce unpredictable results.} (P4)),
and specific calculations (e.g., \usrfb{In our domain, it struggles to compute month-over-month growth rates.} (P5)).
The lack of explicit transmission of this implicit knowledge often leads to errors in the generated SQL.

Faced with repeated failures and misunderstandings by LLMs, participants (7/10) tried to explicitly specify the knowledge by providing more detailed and procedural prompts, but found this approach burdensome.
As P7 reflected, \usrfb{To get the AI to understand what I want, I end up specifying every detail like conditions, fields and joins in natural language.
Finally, I spend more time writing natural language than I would have spent just writing the SQL myself.}
Furthermore, several participants (P1, P3, P8, P10) pointed out that it was error-prone to articulate such knowledge in a clear and structured prompt.
\usrfb{The longer your prompt is, the more likely it is to contain inaccuracies or mistakes.} (P8)
This frustration led several participants (P4, P6) to imagine an ideal system that would be aware of task-specific context.
For instance, \usrfb{It would be better if it understood the project I was working on.
I would just tell it what data I wanted without further explanations.} (P6)

\textbf{Challenges in interpreting implicit knowledge (LLM $\rightarrow$ human)}.
In the \textit{code review} stage, participants (8/10) faced persistent difficulties in making sense of the underlying knowledge behind the LLM-generated SQL scripts,
not only due to poor readability of the code (P6, P7), but more commonly because of the lack of clear alignment between their high-level requirements and the low-level code fragments (P1, P3, P6, P7, P9).
As P9 summarized, \usrfb{SQL syntax itself isn’t hard to understand -- what’s hard is figuring out which of my requirements it meets and which it doesn’t.}
This lack of alignment meant participants struggled to infer what knowledge or assumptions the LLM had made in generating specific parts of the script.
As a result, it became burdensome to exactly locate the code fragments failing to meet their requirements, making subsequent \textit{code refinement} difficult.
To find the error, participants had to repeatedly modify the script and review the execution results of individual subqueries or clauses (P2, P7, P8, P10).
As P10 commented, \usrfb{It takes much effort to find which subquery is wrong, or which step causes the problem.}
Even if participants located the erroneous code fragments, they found it imprecise to edit the code using natural language in the \textit{code refinement} stage (8/10).

\subsection{User Requirements}
\label{ssec:user-requirements}

We identified three requirements for LLM-driven NL-to-SQL tools in SQL authoring based on our findings.
In the remaining parts of the paper, \textbf{R1}-\textbf{R3} denote the user requirements:

\DIFdel{\textbf{R1. Capturing the implicit knowledge for intelligent code suggestions.}
In our interview study, participants preferred to supplement contextual information in their prompts.
However, they expected that the model understood their ``common sense'', or \textit{implicit knowledge}, like methods of computation and dataset-specific conventions.
To achieve more intelligent code suggestions, there is a strong need for LLM-driven NL-to-SQL tools to automatically utilize this implicit knowledge behind user instructions.}

\DIFadd{\textbf{R1. Conveying implicit knowledge.}
In our interview study, participants consistently needed to provide rich contextual information, such as norms and practices of querying a dataset, in order for LLMs to understand their query intentions.
However, the volume and granularity of such information, or \textit{implicit knowledge}, made it burdensome to convey to the LLMs. 
Users therefore need more efficient ways to externalize implicit knowledge without repeatedly specifying every detail during query authoring.}

\DIFdel{\textbf{R2. Supporting knowledge alignment to facilitate code review.}
In our interview study, participants found it challenging to understand LLM-generated SQL scripts.
This difficulty stems from the lack of alignment between the knowledge possessed by users and the knowledge inferred by the model.
To help identify which aspects of users' knowledge are satisfied or missing, LLM-driven NL-to-SQL tools should explain back to users how the inferred knowledge is reflected in the generated query.}

\DIFadd{\textbf{R2. Verifying the use of implicit knowledge}.
Participants found it difficult to interpret how implicit knowledge was manifested in the generated SQL code.
This difficulty often arose because such knowledge was typically organized or presented in an unstructured way, making it challenging for users to identify and verify relevant logic within complex queries.
Bridging this gap requires mechanisms that help users comprehend the knowledge-code relationships, therefore enabling closer alignment between user knowledge and LLM's output.}

\DIFdel{\textbf{R3. Enabling knowledge-level refinement for effective query improvement.}
In our interview study, participants often struggled to effectively modify LLM-generated SQL scripts.
This difficulty primarily arises because it is challenging to locate the code fragments that are not aligned with their knowledge.
To assist users in effective code refinement, LLM-driven NL-to-SQL tools should allow users to directly adjust the inferred knowledge, and automatically apply these changes to the underlying code.}

\DIFadd{\textbf{R3. Refining the generated queries at the knowledge level}.
Participants often needed to iteratively adjust LLM-generated SQL queries when the results did not fully match their expectations.
However, it was difficult for them to conveniently refine queries, because they had to work at the code level and manually locate the specific fragments that reflected their intended knowledge.
Users need more convenient ways to iterate on queries at the knowledge level, so that they can adjust what the query means without laboriously editing low-level SQL code.}

\section{\textit{Cerebra}}
\label{sec:cerebra}
\DIFdel{In this section, we first provide an overview of \mbox{\mytoolname}'s workflow (Section~\mbox{\ref{ssec:cerebra-overview}}).
Then we introduce the definition of implicit knowledge and delineate a classification based on where such knowledge appears within SQL statements (Section~\mbox{\ref{ssec:implicit-knowledge}}).
Following the SQL authoring workflow observed in Section~\mbox{\ref{sssec:how-work-with-nl2sql}}, we dive into how \mbox{\mytoolname} retrieves such knowledge to augment code generation (Section~\mbox{\ref{ssec:retrieve-knowledge}}), presents knowledge for effective code review (Section~\mbox{\ref{ssec:code-review}}), and supports iterative refinement of knowledge to improve scripts (Section~\mbox{\ref{ssec:refine-query}}).}\DIFadd{In this section, we begin with an overview of \mytoolname's workflow, then define and classify implicit knowledge in SQL authoring.
We further describe how \mytoolname\ retrieves and presents this knowledge to support code generation, review, and iterative refinement.}

\subsection{Overview}
\label{ssec:cerebra-overview}

\begin{figure*}[ht]
  \centering
  \includegraphics[width=\linewidth]{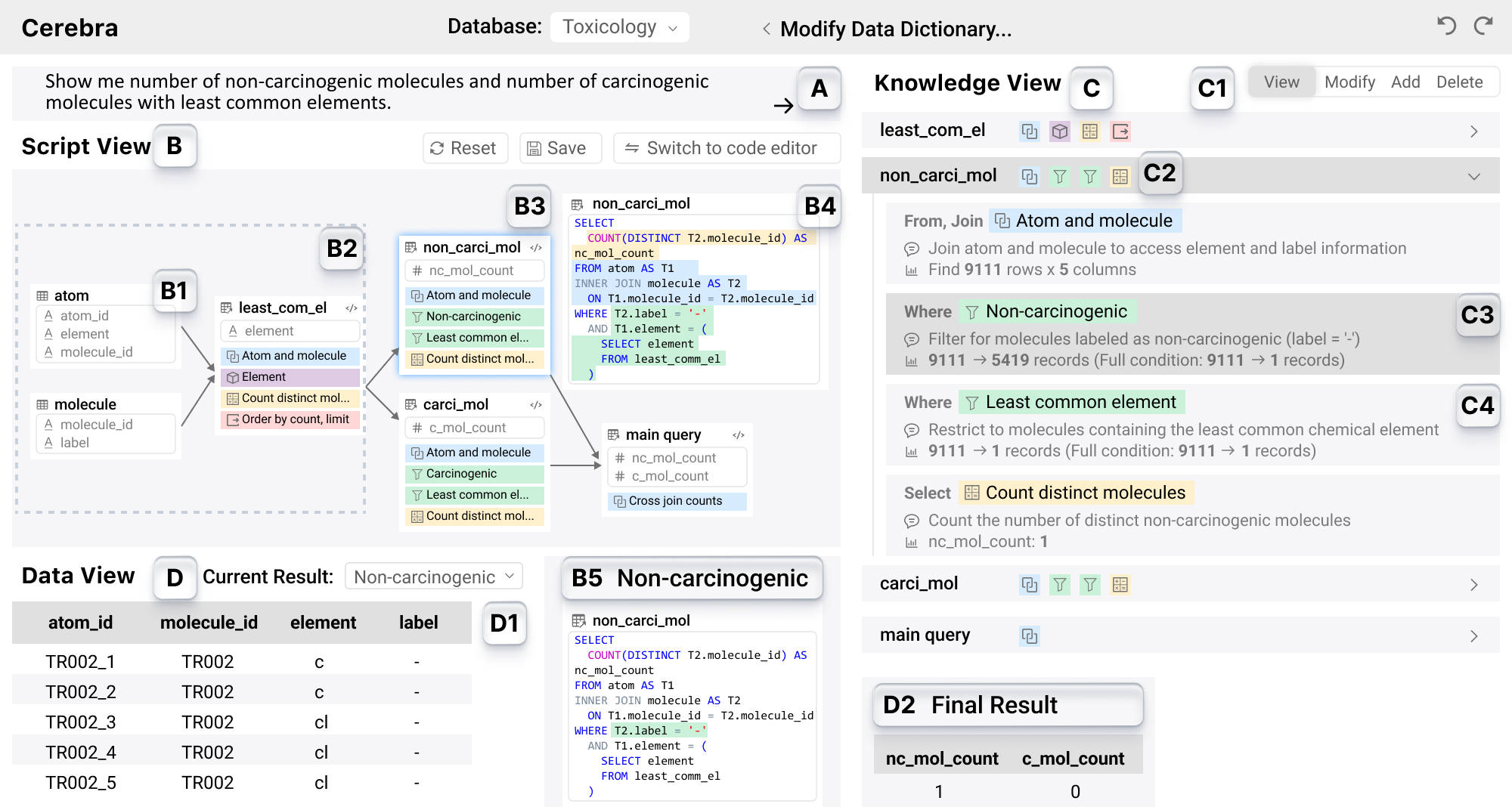}
  \caption{The interface of \mytoolname. A) The input box where users enter their natural language queries. B) The Script View, which visualizes the structure of the generated SQL using a data flow diagram. This view displays input database tables (B1) and subqueries (B3), with links between nodes indicating dependencies (B2). Clicking on a subquery (B3) reveals its corresponding code, with highlights for relevant code fragments (B4). C) The Knowledge View listing implicit knowledge items (e.g., ``non-carcinogenic'' (C3) in the ``non\_carci\_mol'' subquery (C2)). Each knowledge item (C3) is linked to the corresponding code fragments in the subquery (B5). Users can iteratively refine their queries by modifying, adding, or deleting knowledge items (C1). D) The Data View, which displays the intermediate results (D1) of code fragments corresponding to specific knowledge items (C3). It can also display execution results of subqueries, such as the final results of the whole query (D2).}
  \Description{The interface of Cerebra contains four main sections: at the top left (A), an input box where users enter queries; below that (B), the Script View, presenting a data flow diagram with database tables and subqueries as nodes, connected by arrows to indicate data dependencies, and showing the SQL code for the selected subquery with relevant parts highlighted; at the top right (C), the Knowledge View, which lists extracted knowledge items such as ``Non-carcinogenic'' and ``Least common element'' with options to view, modify, add, or delete items and links back to the code; and at the bottom (D), the Data View, displaying a table of intermediate query results for molecules labeled non-carcinogenic, as well as the final query results for the number of non-carcinogenic and carcinogenic molecules with least common elements.}
  \label{fig:interface-init-author}
\end{figure*}

\mytoolname\ is an interactive NL-to-SQL tool that aligns implicit knowledge between users and LLMs during SQL authoring.
It supports SQLite~\cite{sqlite}, though it can be generalized to other SQL dialects.

\subsubsection{System Workflow}
\DIFadd{In our preliminary study (Section~\ref{sssec:how-work-with-nl2sql}), we observed that participants preferred reusing historical scripts during prompt formulation to enhance the model's code generation.
Drawing inspiration from this practice, we incorporated a reuse mechanism directly into \mytoolname's workflow to make the implicit knowledge accessible and facilitate intelligent code suggestions (\textbf{R1}).
As shown in \autoref{fig:cerebra-workflow}, \mytoolname\ adopts a two-stage workflow.
In the offline stage, the system establishes the knowledge base for subsequent query generation.
Specifically, it generates natural language descriptions for database schemas, parses the user's historical SQL scripts into code fragments, and extracts knowledge in the form of natural language.
In the online stage, when a user submits a natural language query, \mytoolname\ retrieves relevant knowledge to augment the LLM prompt.
It then generates the SQL script and structures the output to support the decomposition of complex queries for user review.}

\subsubsection{User Interface}
\DIFadd{The interface (\autoref{fig:interface-init-author}) is designed to support implicit knowledge sensemaking (\textbf{R2}) and knowledge-level query refinement (\textbf{R3}) by visually decomposing a complex request into components.
The Input Box (\autoref{fig:interface-init-author}~A) accepts natural language instructions.
The Script View (\autoref{fig:interface-init-author}~B) visualizes the generated SQL as a data flow diagram, allowing users to trace dependencies between subqueries rather than reading raw code linearly.
The Knowledge View (\autoref{fig:interface-init-author}~C) serves as the core component for alignment.
It translates code fragments into natural language ``knowledge items'', allowing users to verify the knowledge used in this query without parsing complex syntax.
Finally, the Data View (\autoref{fig:interface-init-author}~D) facilitates validation by displaying execution results.
These views are tightly coordinated, such that selecting a component in the diagram highlights the corresponding knowledge item and reveals the intermediate data.
This enables users to iteratively refine the query by focusing on specific components rather than the entire script.}

\DIFdel{As is shown in \mbox{\autoref{fig:interface-init-author}}, the interface of \mbox{\mytoolname} primarily consists of four components:
The input box (\mbox{\autoref{fig:interface-init-author}}~A) where users can enter their natural language instructions,
the Script View (\mbox{\autoref{fig:interface-init-author}}~B) using a data flow diagram to illustrate the dependencies of subqueries,
the Knowledge View (\mbox{\autoref{fig:interface-init-author}}~C) showing the knowledge items,
the Data View (\mbox{\autoref{fig:interface-init-author}}~D) providing the intermediate execution results.
By default, the Data View displays the execution results of the whole query (\mbox{\autoref{fig:interface-init-author}}~D2).
When users click a subquery in the Script View (\mbox{\autoref{fig:interface-init-author}}~B3), \mbox{\mytoolname} will automatically open a code panel beside the subquery (\mbox{\autoref{fig:interface-init-author}}~B4), unfold the inferred knowledge in the Knowledge View (\mbox{\autoref{fig:interface-init-author}}~C2), and display the execution result of the subquery in the Data View.
Users can also click an item in the knowledge View (\mbox{\autoref{fig:interface-init-author}}~C3) to further inspect the corresponding code fragments (\mbox{\autoref{fig:interface-init-author}}~B5) and the intermediate results at this step (\mbox{\autoref{fig:interface-init-author}}~D1).
If the generated script is unsatisfactory, users can switch to ``modify'', ``add'' or ``delete'' mode to update the knowledge (\mbox{\autoref{fig:interface-init-author}}~C1) to iteratively refine the query.}

\DIFdel{In our preliminary study, we observed that participants preferred to reuse historical scripts to enhance code generation by the model during the prompt formulation stage (Section~\mbox{\ref{sssec:how-work-with-nl2sql}}).
Drawing inspiration from participants’ practice, we incorporated this reuse mechanism into the workflow of \mbox{\mytoolname}, to better capture the implicit knowledge for intelligent code suggestions (\textbf{R1}).
As is shown in \mbox{\autoref{fig:cerebra-workflow}}, \mbox{\mytoolname} adopts a two-stage workflow: offline and online.
In the offline stage, to achieve effective code reuse, \mbox{\mytoolname} first takes the database schemas and sample values as input (\mbox{\autoref{fig:cerebra-workflow}} A1) to generate natural language descriptions of each column (\mbox{\autoref{fig:cerebra-workflow}} A2).
It then parses the user-written historical scripts into code fragments (\mbox{\autoref{fig:cerebra-workflow}} A3).
Next, it uses the column descriptions and code fragments to extract knowledge in the form of natural language according to the five knowledge types (\mbox{\autoref{fig:cerebra-workflow}} A4), which will be discussed in Section~\mbox{\ref{sssec:types-implicit-knowledge}}.
The extracted knowledge will be stored and embedded.
In the online stage, as users input a natural language instruction (\mbox{\autoref{fig:cerebra-workflow}} B1), \mbox{\mytoolname} automatically retrieves knowledge relevant to user instructions, generates the SQL script (\mbox{\autoref{fig:cerebra-workflow}} B2), and presents the inferred knowledge items in the Knowledge View for further review and refinement (\mbox{\autoref{fig:cerebra-workflow}} B3).}

\subsection{Implicit Knowledge}
\label{ssec:implicit-knowledge}

\begin{figure*}[ht]
  \centering
  \includegraphics[width=\linewidth]{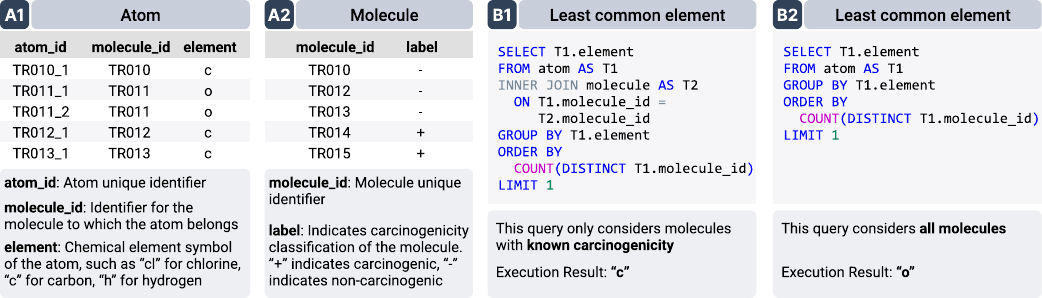}
  \caption{An example of task-specific computations, which represent a form of implicit knowledge. Suppose data practitioners define the ``least common element'' as computed in B1 when working with the Toxicology database using the table ``atom'' (A1) and ``molecule'' (A2). If the model does not capture this implicit knowledge, it may produce incorrect results such as B2.}
  \Description{Figure showing an example of task-specific computations using tables and SQL queries with the Toxicology database. On the left, two tables are displayed: ``Atom'' (A1) with columns atom\_id, molecule\_id, and element, listing chemical elements for several atoms, and ``Molecule'' (A2) with columns molecule\_id and label, where the label indicates carcinogenicity (``+'' for carcinogenic, ``-'' for non-carcinogenic). Below each table is a brief explanation of the columns. To the right, two panels describe different SQL queries for finding the ``least common element.'' Panel B1 shows a query that considers only molecules with known carcinogenicity, joining the two tables and grouping by element, resulting in ``c'' as the least common element. Panel B2 shows a query that does not filter for carcinogenicity and instead considers all molecules, resulting in ``o'' as the least common element. Each query's logic and the difference in execution results are explained below the SQL code, illustrating how omitting implicit knowledge about filtering can lead to incorrect results.}
  \label{fig:different-computation}
\end{figure*}

Implicit knowledge in the NL-to-SQL process refers to the specifications that are not explicitly provided in users’ natural language instructions.
Typical implicit knowledge includes dataset-specific conventions and task-specific computations, which users tend to assume are understood by the model.
\DIFdel{In this subsection, we first illustrate implicit knowledge through two examples (Section~\mbox{\ref{sssec:example-implicit-knowledge}}).
We then categorize the types of implicit knowledge according to where they manifest in SQL queries (Section~\mbox{\ref{sssec:types-implicit-knowledge}}),
and later leverage this categorization to extract implicit knowledge from users' past SQL scripts (Section~\mbox{\ref{ssec:retrieve-knowledge}}).}
\DIFadd{Throughout the rest of this paper, we use the term \textit{knowledge item} to refer to a piece of implicit knowledge, such as a filtering condition or calculation formula.
Each knowledge item is associated with a specific code fragment or subquery in the SQL script.}

\subsubsection{Implicit Knowledge Examples}
\label{sssec:example-implicit-knowledge}

\sloppy

To illustrate how implicit knowledge manifests in practice, we present two examples from the Toxicology database (\autoref{fig:toxicology-schema}) mentioned in Section~\ref{sec:intro}.
We focus on implicit knowledge related to the dataset-specific conventions and task-specific computations.

\fussy

\textbf{Example of dataset-specific conventions.}
Assume a data practitioner in chemistry is using the Toxicology database.
She knows that ``carcinogenic molecules'' refers to the molecules with the ``+'' value in the ``label'' column (\autoref{fig:toxicology-schema} B).
Here, the mapping from ``carcinogenic'' to the ``label'' column with value ``+'' is a form of implicit knowledge, since it is difficult to directly infer the meaning of this column just from its name, ``label''.
However, she often assumes that the NL-to-SQL model possesses such knowledge, and thus uses terms like ``carcinogenic'' directly in her natural language instructions without explicitly specifying the corresponding column or value in the database.

\textbf{Example of task-specific computations.}
Another type of implicit knowledge arises from conventions in how certain computations are performed.
For example, when a data practitioner is querying the tables ``atom'' (\autoref{fig:different-computation} A1) and ``molecule'' (\autoref{fig:different-computation} A2),
she typically interprets the ``least common element'' as the least common element among molecules with known carcinogenicity (\autoref{fig:different-computation} B1).
In contrast, the model may default to computing the ``least common element'' across all molecules in the database, regardless of whether their carcinogenicity is known (\autoref{fig:different-computation} B2).
This difference indicates that task-specific computations are often implicitly assumed by data practitioners,
but may not be explicitly stated in natural language, thus require the model to possess such implicit knowledge for correct interpretation.

\subsubsection{Implicit Knowledge Types}
\label{sssec:types-implicit-knowledge}

We categorize implicit knowledge according to where it appears in SQL queries based on the grammar of SQLite~\cite{sqlite},
a lightweight relational database that is widely adopted in various NL-to-SQL datasets~\cite{spider, bird, spider2, drspider}.
We identify five types of implicit knowledge: calculation, condition, relation, dimension, and output.
Below, we describe each type in detail.

\textbf{Calculation (Expressions in \texttt{SELECT}/\texttt{ORDER BY} clauses).}
Calculation knowledge involves assumptions about how attributes are computed, which are often reflected in specific formulas or the use of functions.
For example, when a user asks for the ``click-through rate'' (CTR), the precise calculation varies between teams or organizations,
as the term can refer to clicks divided by page views ($CTR = clicks\ /\ page\ views$) or clicks divided by unique users ($CTR = clicks\ /\ unique\ users$), or other more complex formulas, even though the same term is used.

\textbf{Condition (Expressions in \texttt{WHERE}/\texttt{HAVING} clauses).}
Condition knowledge captures underspecified constraints that users expect, often conveyed through modifiers or contextual cues.
For example, a query like ``\textit{calculate the number of molecules containing non-bonding elements}'' leaves the definition of ``\textit{non-bonding elements}'' unspecified.
What counts as a non-bonding element may differ across teams or organizations, similar to calculation knowledge.

\textbf{Relation (\texttt{FROM}-\texttt{JOIN} clauses).}
Relation knowledge refers to understanding which tables should be joined and how they are connected.
For instance, as illustrated in the task-specific computations example (Section~\ref{sssec:example-implicit-knowledge}), when a data practitioner queries ``\textit{the least common element}'',
this requires joining the ``atom'' and ``molecule'' tables (\autoref{fig:different-computation} B1), even though the join logic is not explicitly specified in the natural language instruction.

\textbf{Dimension (\texttt{GROUP BY} clauses).}
Dimension knowledge relates to the granularity at which results are aggregated.
For example, when a data practitioner asks for ``\textit{the average order value}'',
she does not specify whether the average should be calculated across all orders, per region, or by a combination of dimensions such as region and customer type.

\textbf{Output (\texttt{ORDER BY}/\texttt{LIMIT}/\texttt{DISTINCT} clauses).}
Output knowledge involves how results should be presented, such as ordering, limiting, or deduplication.
Similar to the dimension knowledge, users may not state these requirements explicitly, but there are often conventions or expectations for the output data.

\subsection{Usage Scenario}
\begin{figure*}[ht]
  \centering
  \includegraphics[width=\linewidth]{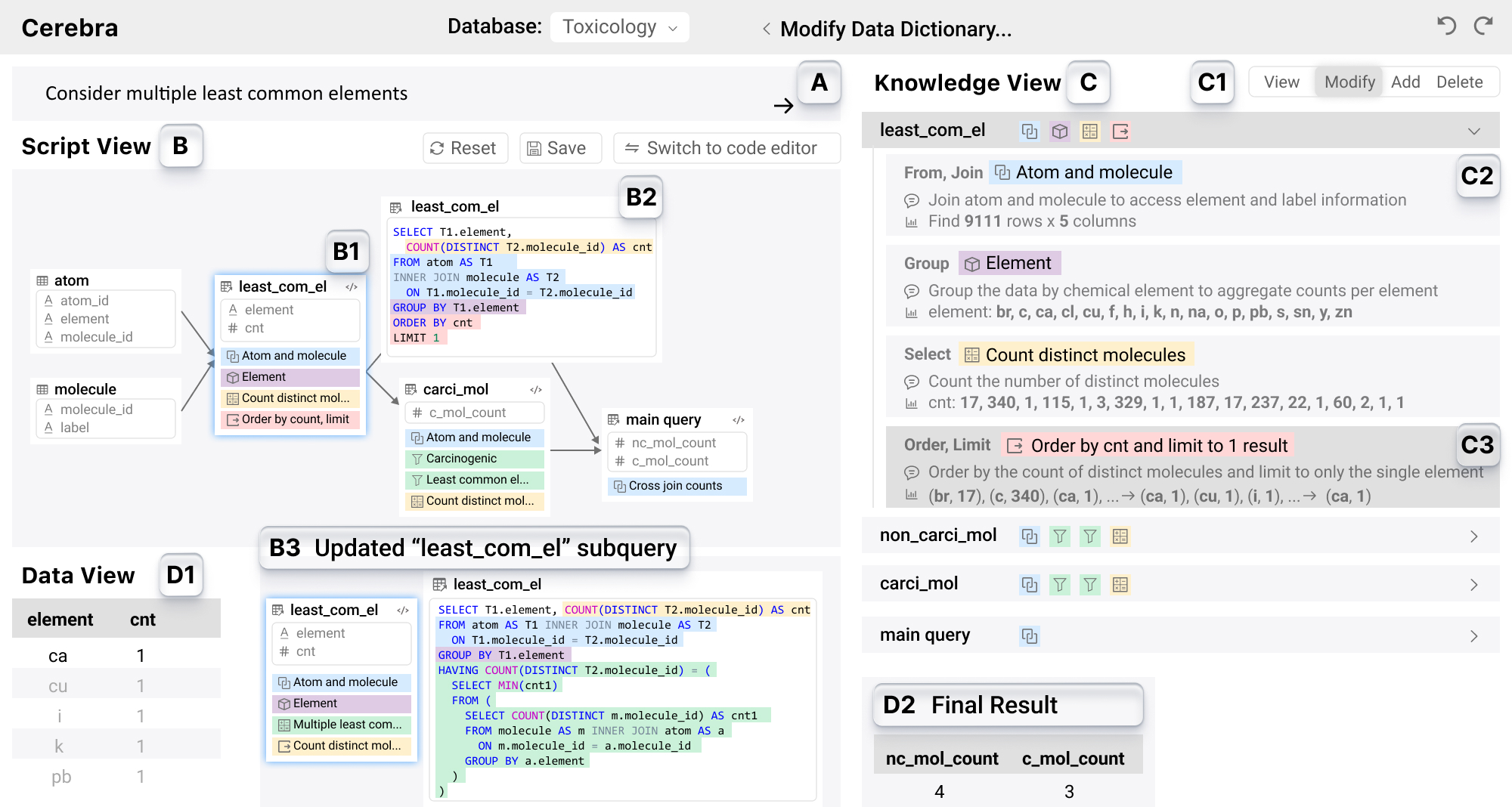}
  \caption{The usage scenario of refining the generated scripts. A) The input box, where users can enter natural language instructions to improve the query. B) The Script View, which displays the subquery that queries the least common elements (B1) and its corresponding code (B2). The refined script is shown in B3. C) The Knowledge View, which lists implicit knowledge items (e.g., C2, C3). Users can switch to ``modify'' mode (C1) to update the knowledge item (C3). D) The Data View, which shows the intermediate results (D1) of the knowledge item (C3). After refinement, the result of the whole query is shown in D2.}
  \Description{The Cerebra interface shows the scenario of refining generated scripts for querying a toxicology database, with four main panels: at the top left, an input box (A) for entering natural language instructions; below it, the Script View (B) displays a flowchart of linked tables and subqueries, including a highlighted "least\_com\_el" subquery (B1) with its corresponding SQL code (B2) and an updated version (B3); to the right, the Knowledge View (C) summarizes the query's implicit knowledge as editable steps, such as joining tables (C2), grouping by element, and ordering results (C3), with a "Modify" button (C1) for switching modes; at the bottom left, the Data View (D) lists intermediate results as a table of element-count pairs (D1); at the bottom right, the final query results (D2) are shown as counts of non-carcinogenic and carcinogenic molecules.}
  \label{fig:refine-script}
\end{figure*}

Assume that Emma, a data practitioner in Chemistry, was working on a toxicology database project.
She had authored many SQL scripts with the aid of LLM-driven NL-to-SQL tools to answer a wide range of questions.
However, Emma noticed that the underlying model often generated incorrect SQL code, leading to considerable effort in carefully reviewing the generated scripts.
Even after pinpointing the issues, Emma still needed to write lengthy prompts to repeatedly explain the dataset-specific conventions and clarify the task-specific computations.
Seeking better coding assistance, she turned to \mytoolname.
Emma set up her workspace by connecting to the database and importing her historical SQL scripts, enabling \mytoolname\ to extract knowledge automatically.

To start the query, Emma entered a natural language instruction in the input box (\autoref{fig:interface-init-author} A):
``\textit{Show me number of non-carcinogenic molecules and number of carcinogenic molecules with least common elements.}''
Based on this instruction, \mytoolname\ automatically retrieved relevant knowledge from Emma's past queries and generated a SQL script in the Script View (\autoref{fig:interface-init-author}~B).
Emma noticed a significant improvement compared to other NL-to-SQL tools she had used previously.
\mytoolname\ had automatically joined the ``atom'' and ``molecule'' tables (\autoref{fig:interface-init-author}~B2) to compute the least common elements in the ``least\_com\_el'' subquery, saving her effort of repeated clarification.

However, Emma was skeptical about the result (\autoref{fig:interface-init-author}~D2), as she believed there should be more than one molecule that met the criteria, so she first clicked the ``non\_carci\_mol'' subquery (\autoref{fig:interface-init-author}~B3) to check the inferred knowledge.
In the Knowledge View, she immediately verified that \mytoolname\ had correctly interpreted ``Non-carcinogenic'' as filtering molecules with label = ``-'' (\autoref{fig:interface-init-author}~C3).
Besides, she noticed that the ``Least common element'' (\autoref{fig:interface-init-author}~C4) only retained a single record.
Realizing that this might be due to an error in the previous ``least\_com\_el'' subquery, she clicked the subquery (\autoref{fig:refine-script}~B1) and identified an output knowledge item (\autoref{fig:refine-script}~C3) that was restricting the result to one record (\autoref{fig:refine-script}~D1).
Emma identified the cause:
she had previously authored a script to query ``one of the least common elements,'' causing \mytoolname\ to automatically incorporate a \texttt{LIMIT 1} clause in the current SQL script.

To adapt the query to her current task, Emma switched to the ``modify'' mode (\autoref{fig:refine-script}~C1).
She selected the problematic knowledge item (\autoref{fig:refine-script}~C3), and entered the instruction ``\textit{Consider multiple least common elements}'' in the input box (\autoref{fig:refine-script}~A).
\mytoolname\ responded by replacing the output knowledge in red with a new condition knowledge in green (\autoref{fig:refine-script}~B3), and regenerated the corresponding script of the subquery.
Emma observed that the final result changed from (1, 0) in \autoref{fig:interface-init-author}~D2 to (4, 3) in \autoref{fig:refine-script}~D2, consistent with her expectations.
By reviewing the script and intermediate results, Emma verified the correctness of the outcome and completed her query authoring task.

\subsection{Retrieving Knowledge for Intelligent Code Suggestions}
\label{ssec:retrieve-knowledge}

To offer intelligent code suggestions (\textbf{R1}), \mytoolname\ retrieves knowledge and injects it in prompts during the code generation process.
\DIFdel{In this subsection, we first describe how the knowledge extraction process is set up (Section~\mbox{\ref{sssec:setup-know-extract}}).
We then detail the process of parsing code fragments and mapping them to the five implicit knowledge types (Section~\mbox{\ref{sssec:extract-implicit-knowledge}}).
Furthermore, we explain how relevant knowledge is retrieved and incorporated into code generation (Section~\mbox{\ref{sssec:inject-knowledge}}).}

\subsubsection{Setting Up Knowledge Extraction Process}
\label{sssec:setup-know-extract}

\begin{figure*}[ht]
  \centering
  \includegraphics[width=\linewidth]{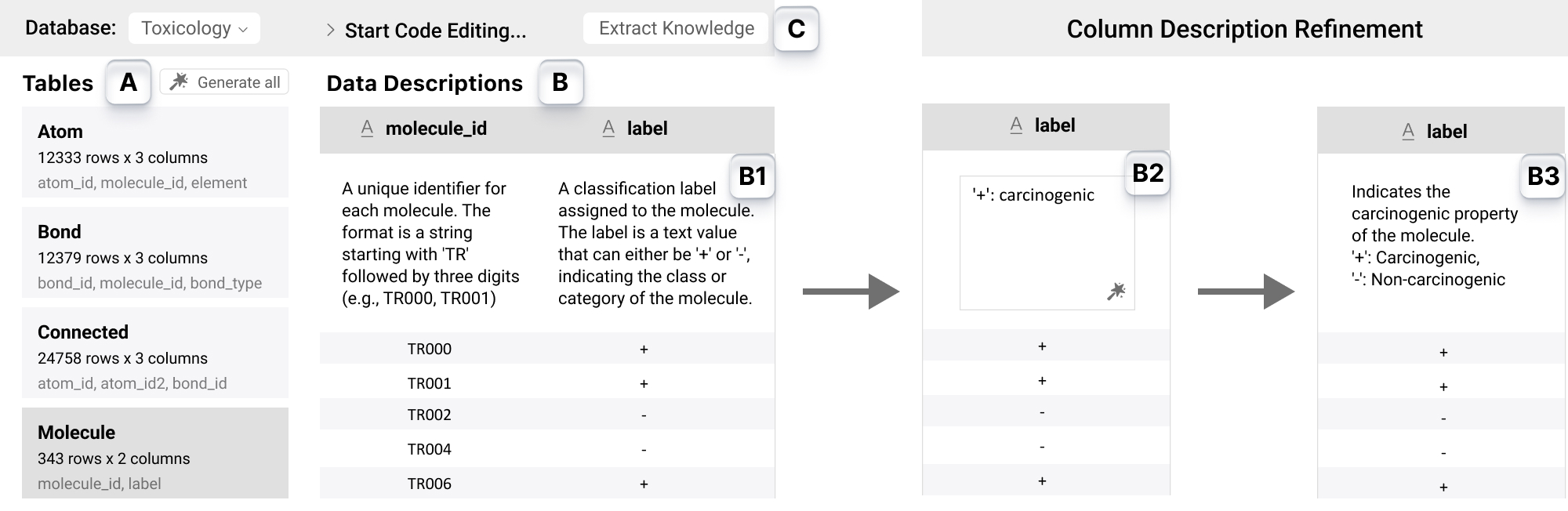}
  \caption{The data dictionary interface of \mytoolname. A) The Table View lists basic metadata for each database table, including shapes and column names. B) The Data Descriptions View displays sample values along with generated descriptions for each column. Users can edit the column descriptions by double-clicking on the descriptive text (B1) and entering partial descriptions (B2). \mytoolname\ will generate the remainder of the description based on the unique values and format of the column (B3).}
  \Description{Cerebra data dictionary interface showing two main sections: on the left, the Table View lists database tables such as Atom, Bond, Connected, and Molecule, each with row and column counts and sample column names; on the right, the Data Descriptions View displays metadata and sample values for the selected table's columns, including generated descriptions for columns like molecule\_id and label, with editable text fields where users can modify or partially enter descriptions, and an example in which the label column's description is incrementally edited to indicate that '+' means carcinogenic and '-' means non-carcinogenic, illustrating how the interface supports interactive editing and automatic completion of data descriptions.}
  \label{fig:data-dict}
\end{figure*}

\mytoolname\ extracts knowledge leveraging users' past scripts.
However, it is not a straightforward process since the accompanying code documentation such as data dictionaries is usually unavailable~\cite{doc-db-usage, documentation-matters}, which makes it difficult to interpret the high-level semantics behind the scripts.
To address this challenge, we designed an auxiliary data dictionary view in \mytoolname\ (\autoref{fig:data-dict}).
When users first import their database and historical scripts, they can click ``Modify Data Dictionary...'' on the top navigation bar (\autoref{fig:interface-init-author}) to switch to the data dictionary view,
where they are able to interactively edit the column descriptions to supplement it with dataset-specific conventions.

The data dictionary view consists of a table list (\autoref{fig:data-dict} A) which lists all data tables in the database and a data description view (\autoref{fig:data-dict} B) offering column descriptions as well as sample data.
\mytoolname\ generates initial column descriptions according to the sample data.
If users find the description of a certain column is vague (\autoref{fig:data-dict} B1), they can double-click the descriptive text and input their partial descriptions (\autoref{fig:data-dict} B2), and let \mytoolname\ complete the rest of the description (\autoref{fig:data-dict} B3).
Once users fix the data dictionary, they can click the button (\autoref{fig:data-dict} C) to extract knowledge based on the dictionary and past scripts (Section~\ref{sssec:extract-implicit-knowledge}).

While interactive editing of column descriptions lowers users' effort in building a data dictionary from scratch, it can still be tedious when many columns have unclear names like the name ``label'' (\autoref{fig:data-dict} B1).
To mitigate the issue, \mytoolname\ also considers column aliases in users' past queries (e.g., \texttt{SELECT label AS flag\_carcinogenic}), as these aliases often capture users' own interpretations.
\DIFadd{These aliases are automatically parsed and used internally to enhance the quality of suggested column descriptions.}
By leveraging such aliases alongside sample data, \mytoolname\ can better suggest meaningful column descriptions and further reduce the effort of manual annotation.

\subsubsection{Extracting Knowledge from Historical Code Fragments}
\label{sssec:extract-implicit-knowledge}

Directly prompting LLMs to extract knowledge from historical scripts often leads to syntactic errors in the identified code fragments, and the extracted fragments cannot be always executable to obtain the intermediate results.
To address these issues, \mytoolname\ first employs a rule-based approach to parse historical scripts into code fragments,
before it leverages LLMs to interpret each fragment to obtain the knowledge in natural language form.

\mytoolname\ decomposes historical scripts into well-formed code fragments that correspond to the five types of implicit knowledge (Section~\ref{sssec:types-implicit-knowledge}) through the Abstract Syntax Tree (AST).
For calculation knowledge, expressions are extracted by splitting the \texttt{SELECT} or \texttt{ORDER BY} clause at commas, treating each formula as a standalone calculation fragment, while preserving the integrity of nested formulas.
For condition knowledge, compound logical conditions in \texttt{WHERE} or \texttt{HAVING} clauses are further decomposed into atomic conditions, ensuring that each simple predicate is extracted as an individual fragment.
For the remaining knowledge types (dimension, output, and relation), the corresponding \texttt{GROUP BY}, \texttt{ORDER BY}/\texttt{LIMIT}/\texttt{DISTINCT}, and \texttt{FROM}/\texttt{JOIN} clauses are each extracted as a whole, preserving their original structure.
Detailed parsing rules can be found in the supplementary material.

After parsing, \mytoolname\ leverages LLMs to interpret each historical query by first generating a natural language description of the whole script using the data dictionary (\autoref{fig:data-dict}) for context.
For example, for the query in \autoref{fig:different-computation}~B1, the script-level description is:
``\textit{Find the least common element.}''
Then, \mytoolname\ prompts the LLM to produce descriptions for each individual code fragment.
For instance, the fragment \texttt{FROM atom AS T1 INNER JOIN molecule AS T2} is described as:
``\textit{Join atoms and molecules to filter molecules with known carcinogenicity.}''
Both script-level and fragment-level descriptions are embedded for retrieval, enabling \mytoolname\ to better match user instructions to relevant knowledge.

\subsubsection{Injecting Knowledge into Prompts}
\label{sssec:inject-knowledge}

When users input a natural language instruction (\autoref{fig:interface-init-author} A), \mytoolname\ retrieves both script-level and fragment-level knowledge before injecting the knowledge into prompts to generate SQL scripts.
Initially, it matches the user instruction against the embedded script-level descriptions of historical queries, using cosine similarity to measure semantic relevance.
Given that the user instructions often do not correspond exactly to any single past script,
\mytoolname\ then decomposes the user’s instruction into multiple keywords with the help of the LLM,
and matches each keyword with the embedded fragment-level descriptions.
For example, given an instruction such as ``\textit{Show me number of non-carcinogenic molecules with non-bonding element}'', \mytoolname\ extracts keywords like ``\textit{non-carcinogenic molecules}'' and ``\textit{non-bonding element}'', and then retrieves relevant SQL fragments corresponding to these keywords.
Additionally, \mytoolname\ presents all retrieved scripts and associated knowledge to the LLM for filtering and re-ranking, prioritizing examples most relevant to the current task.
Finally, \mytoolname\ uses the re-ranked scripts, associated knowledge, and the data dictionary as input for the LLM to generate the SQL code.

\subsection{Making Implicit Knowledge Explicit for Effective Code Review}
\label{ssec:code-review}

To address the challenge of knowledge-code misalignment identified in our preliminary study, \mytoolname\ is designed to facilitate code review by presenting both the code structure and the knowledge inferred during code generation (\textbf{R2}).
This enables users to easily inspect which aspects of their knowledge are satisfied or missing in the generated query. 
\DIFdel{Below, we describe the two main components: the Script View and the Knowledge View.}

\subsubsection{Visualizing the Data Flow of SQL Queries in the Script View}
\label{sssec:script-view}

The Script View (\autoref{fig:interface-init-author}~B) provides users with an overview of the entire SQL script.
It uses a data flow diagram to visualize the table-level lineage of the generated SQL script, including input tables (e.g., \autoref{fig:interface-init-author}~B1), subqueries (e.g., \autoref{fig:interface-init-author}~B3), and their dependencies.
Each subquery is represented as a card displaying key metadata such as subquery names, output columns, column types, and a list of knowledge items.
Clicking on a subquery card (\autoref{fig:interface-init-author}~B3) highlights the corresponding SQL code (\autoref{fig:interface-init-author}~B4) and shows the intermediate execution results.

To support the inspection of intermediate execution results, \mytoolname\ computes a dependency graph among subqueries based on the ASTs of the SQL script.
For each subquery, dependencies are recursively resolved by identifying all referenced tables and Common Table Expressions (CTEs)\DIFdel{, following the algorithm shown in Algorithm~1}.
Therefore, to compute the execution result of a specific subquery, \mytoolname\ will automatically execute all its dependent subqueries in the correct order to avoid syntax errors caused by undefined subqueries.

\subsubsection{Presenting Knowledge in the Knowledge View}
\label{sssec:knowledge-view}

The Knowledge View (\autoref{fig:interface-init-author}~C) presents a structured list of the knowledge items extracted from the SQL script, organized by subqueries and clauses.
All knowledge items are grouped in a flat, two-level hierarchy:
the outer level corresponds to subquery names (\autoref{fig:interface-init-author}~C2), while the inner level lists individual knowledge items (e.g., \autoref{fig:interface-init-author} C3) associated with specific SQL code fragments, following the execution order.
While this flat, two-level design inevitably omits some hierarchical information of knowledge items (e.g., two dependent knowledge items in different subqueries), it simplifies reading without being overwhelmed by deeply nested SQL structures.
Besides, users can still inspect the hierarchical dependencies by referring to the Script View if needed.

Each knowledge item corresponds to one of the five implicit knowledge types (see Section~\ref{sssec:types-implicit-knowledge}) and is presented with a two-line display:
the first line is a code fragment-level natural language description similar to that in Section~\ref{sssec:extract-implicit-knowledge}, while the second line presents relevant metadata of execution results of the code fragment for further validation.
Specifically, for the code fragment corresponding to calculation knowledge, the metadata shows sample values computed by the expression in the \texttt{SELECT} statement;
for condition knowledge, it reports both the number of records filtered by the atomic condition and by the overall composite condition;
for relation knowledge, it displays the number of rows and columns after joining the relevant tables;
for dimension knowledge, it lists the distinct values present in the aggregation dimension;
and for output knowledge, it presents a sample of the final output records.

When users click on a knowledge item (\autoref{fig:interface-init-author} C3), \mytoolname\ displays the complete execution results in the Data View (\autoref{fig:interface-init-author} D1), along with highlighting the corresponding code fragment (\autoref{fig:interface-init-author}~B5).
Since computing intermediate results for code fragments can be time-consuming, \mytoolname\ adopts an asynchronous loading strategy to incrementally present results as they become available, helping shorten user waiting time and improve interactivity.

\subsection{Refining Knowledge for Effective Query Modification}
\label{ssec:refine-query}

While historical script retrieval enables \mytoolname\ to capture the implicit knowledge embedded in users' prior work, we found that code reuse alone is insufficient for supporting users' evolving query requirements.
To better support code editing, \mytoolname\ allows users to incrementally refine the queries by modifying, adding, or deleting (\autoref{fig:interface-init-author}~C1) the knowledge in the Knowledge View (\textbf{R3}).

When modifying, users can select a subquery (e.g., \autoref{fig:interface-init-author} C2) or a specific knowledge item (e.g., \autoref{fig:interface-init-author} C3), and enter an instruction in the input box (\autoref{fig:interface-init-author} A) to update the corresponding code fragments.
To add new knowledge, users can build on an existing subquery or select the entire SQL query (i.e., the main query in \autoref{fig:interface-init-author}) and provide an additional instruction.
This triggers \mytoolname\ to retrieve and integrate relevant knowledge from historical scripts.
For deleting, users simply select knowledge items to remove them from the query.
\DIFdel{For all three editing operations, \mbox{\mytoolname} automatically propagates necessary changes to all dependent subqueries, preventing errors such as undefined fields and maintaining the syntactic correctness of the entire query.}

\DIFadd{To ensure the syntactic correctness of the entire query after any modification, \mbox{\mytoolname} maintains a dependency graph for all subqueries and their associated knowledge items, constructed from the AST of the whole SQL script.
When a user edits a knowledge item, \mbox{\mytoolname} first locates the subquery that incorporates the modified code fragment.
It then recursively identifies all downstream subqueries that depend on the outputs of the affected subquery, following the edges in the dependency graph.
For each downstream subquery, \mbox{\mytoolname} automatically updates the SQL code to reflect changes in referenced fields or aliases.
It then re-executes these subqueries in topological order, ensuring that any intermediate results and final outputs remain consistent with the updated knowledge.}

\subsection{System Implementation}
\label{ssec:sys-impl}

We implemented the frontend of \mytoolname\ using TypeScript in web browsers, in combination with libraries involving Vue.js, Monaco Editor~\cite{monaco-editor}, and dagre~\cite{dagre-library}.
The Monaco Editor was used to display and highlight the source code of subqueries, while dagre was used for automatic layout of the dataflow diagram in the Script View.
The backend of \mytoolname\ was implemented in Python, utilizing SQLGlot~\cite{sqlglot} for SQL parsing, SentenceTransformers~\cite{sbert} to compute word embeddings, and Qwen3~\cite{qwen3} as the underlying LLM.

\section{User Study}
\label{sec:user-study}
To evaluate the effectiveness and usability of \mytoolname, we conducted a user study\footnote{The study has received approval from State Key Lab of CAD\&CG, Zhejiang University.}
where participants completed SQL authoring tasks using \mytoolname\ and a baseline tool.
\CAMadd{We focus on research questions:}

\begin{itemize}
    \item \CAMadd{\textbf{RQ1 (overall experience)}: To what extent does \mytoolname\ improve the NL-to-SQL authoring process?}
    \item \CAMadd{\textbf{RQ2 (steering strategies)}: How do users navigate different steering mechanisms to align the model with their intent?}
    \item \CAMadd{\textbf{RQ3 (mental models)}: How do users understand implicit knowledge and adjust the explicitness of their prompts?}
    \item \CAMadd{\textbf{RQ4 (transparency \& trust)}: How does \mytoolname\ affect users' understanding and trust in the generated SQL?}
\end{itemize}

\subsection{Participants}
\label{ssec:us-participants}

In our user study, we recruited 16 data practitioners (denoted as U1-U16, 10 male and 6 female, $Age_{mean}=25.75$, $Age_{std}=5.57$) through two channels:
11 participants were recruited via a university’s internal forum, and the other 5 were recruited from industry through social media.
Participants had diverse backgrounds, including Cloud Computing, Biology, Chemical Engineering, Gaming, and E-Commerce.
All participants were expert in SQL programming ($Experience_{mean}=3.94$ years, $Experience_{std}=4.02$ years) and routinely authored queries in their work (at least once a week).
They also reported moderate familiarity with LLM-driven NL-to-SQL tools ($M = 4.8$) on a 7-point Likert scale (1 = not at all familiar, 7 = extremely familiar).
\DIFdel{Their detailed demographic information is provided in the supplementary materials.}
\DIFadd{Their detailed demographic information is presented in \autoref{tab:demographics-user} of the appendix.}
All participants gave consent for the recording of both their voices and programming processes.

\subsection{Apparatus and Materials}
\label{ssec:us-apparatus-materials}


\subsubsection{Baseline}
To evaluate the integration of knowledge in \mytoolname, we designed the baseline tool (denoted as Baseline) as an ablation version that removed all features related to knowledge in both code generation and interface design.
Specifically, in code generation, Baseline used the same backend LLM, but generated SQL queries solely based on user instructions, data schemas, and column descriptions, without retrieving knowledge from existing scripts.
In the interface, Baseline omitted all knowledge presentation by replacing the Knowledge View with a simple chat interface, where users could enter natural language instructions and receive generated SQL scripts with textual explanations.
The data flow diagram in the Script View was retained, but knowledge items in the subquery cards and their corresponding highlights in the code were removed.
Baseline enabled users to generate, review, and refine SQL scripts via LLM-driven code generation.

\subsubsection{Datasets}
To design a comparative study, we selected two datasets, denoted as \textit{European Football} and \textit{Toxicology}, on the cross-domain NL-to-SQL dataset, BIRD~\cite{bird}.
Compared to other datasets such as Spider~\cite{spider} and Spider 2.0~\cite{spider2}, BIRD features not only complex and large-scale databases, but also a larger number of SQL scripts per individual database.
This allowed us to better evaluate the effectiveness of knowledge extraction and historical script reuse in \mytoolname. 
Each dataset contains 11 data columns and more than 50 existing SQL scripts.
Both datasets come with predefined database schemas and column descriptions, which were provided to participants directly and did not require any user editing.



\subsubsection{Tasks}
For each dataset, we designed a statistical task that required participants to compose a SQL query.
The final query of each task was designed to be approximately 30 lines long and to include 3 subqueries.
To reduce the impact of initial model output deviations on the experiment, we provided a dedicated initial prompt for each task, following the practice adopted in prior work~\cite{waitgpt}.
Detailed task descriptions, including initial prompts and expected outputs, are provided in the supplementary materials.


\subsection{Procedure}
\label{ssec:us-procedure}
We adopted a counterbalanced mixed design to compare \mytoolname\ and Baseline.
We denoted the two tools as C(erebra) and B(aseline), and the two datasets as E(uropean Football) and T(oxicology).
Participants were divided into four groups to cover all combinations of tools and datasets in the following experimental conditions: [CT, BE], [BE, CT], [BT, CE], [CE, BT],
allowing each participant to experience both tools and mitigating learning effects.

At the start of the study, we informed participants about the relevant background.
We then introduced the first code authoring task, during which they practiced with the assigned tool in a warm-up phase and received a slide deck containing task descriptions, relevant knowledge, and data dictionaries.
After completing the first task, which involved authoring SQL scripts with either \mytoolname\ or Baseline, participants filled out a NASA-TLX~\cite{nasa-tlx} questionnaire,
with each question measured on a 7-point Likert scale, to assess perceived workload.
Subsequently, we switched to the second code authoring task using a different dataset and tool, following the same procedure.
After both code authoring tasks, we conducted a semi-structured interview comprising three parts.
First, we asked participants to compare their experiences with \mytoolname\ and Baseline in terms of code generation, understanding, and refinement.
Second, we encouraged participants to share their suggestions for improving the tool.
Third, we asked participants follow-up questions regarding specific issues that arose during the SQL authoring process.
The entire study took around 80 minutes and each participant received 70 Chinese Yuan as compensation.

\CAMadd{During the study, we recorded all screen activities and audio.
To analyze the interaction patterns, two authors (SQL authoring experience: $> 3\ years$) manually annotated participants' actions from the screen-capture videos.
Disagreements were discussed and resolved in periodic meetings.
When consensus could not be reached, a third author (SQL authoring experience: $> 10\ years$) was consulted to make the final decision.
For the semi-structured interviews, we transcribed all audio recordings and conducted an inductive content analysis~\cite{content-analysis} to identify recurring themes and patterns in participants' feedback.}





\CAMdel{To evaluate whether \mbox{\mytoolname} accelerated the script authoring process, we measured ``task completion time'' of each participant of each task.
Since the raw data of task completion time obey both normal distribution and equal variance, we used ANOVA for statistical tests.
We tested the main effects of tools and tasks, as well as their interaction effects.
We found that the main effects of tools are significant ($p < 0.01$), while the interaction effects and the main effects of tasks are not significant.
According to the result of task completion time (\mbox{\autoref{fig:comptime}}), participants spent less time completing the tasks with \mbox{\mytoolname} than with Baseline, reflecting the overall advantage indicated by the ANOVA.}

\CAMdel{For perceived workload, we analyzed the NASA‑TLX data, as shown in \mbox{\autoref{fig:workload}}.
Since only the Mental dimension met normality and equal‑variance assumptions, we applied a standard ANOVA to Mental and used non‑parametric Aligned Rank Transform (ART) ANOVA for the remaining dimensions.
The analysis revealed significant main effects of tools on the Performance ($p < 0.01$) and Effort ($p < 0.05$) workload scores.
Post‑hoc EMMeans comparisons confirmed that participants reported significantly lower Performance and Effort workload when using \mbox{\mytoolname} than when using Baseline.
All other NASA‑TLX dimensions (Mental, Physical, Temporal, and Frustration) showed no significant effect of tools.
Across all dimensions, neither the main effect of tasks nor the interaction between tools and tasks was significant.}

\begin{figure*}[ht]
  \centering
  \includegraphics[width=\textwidth]{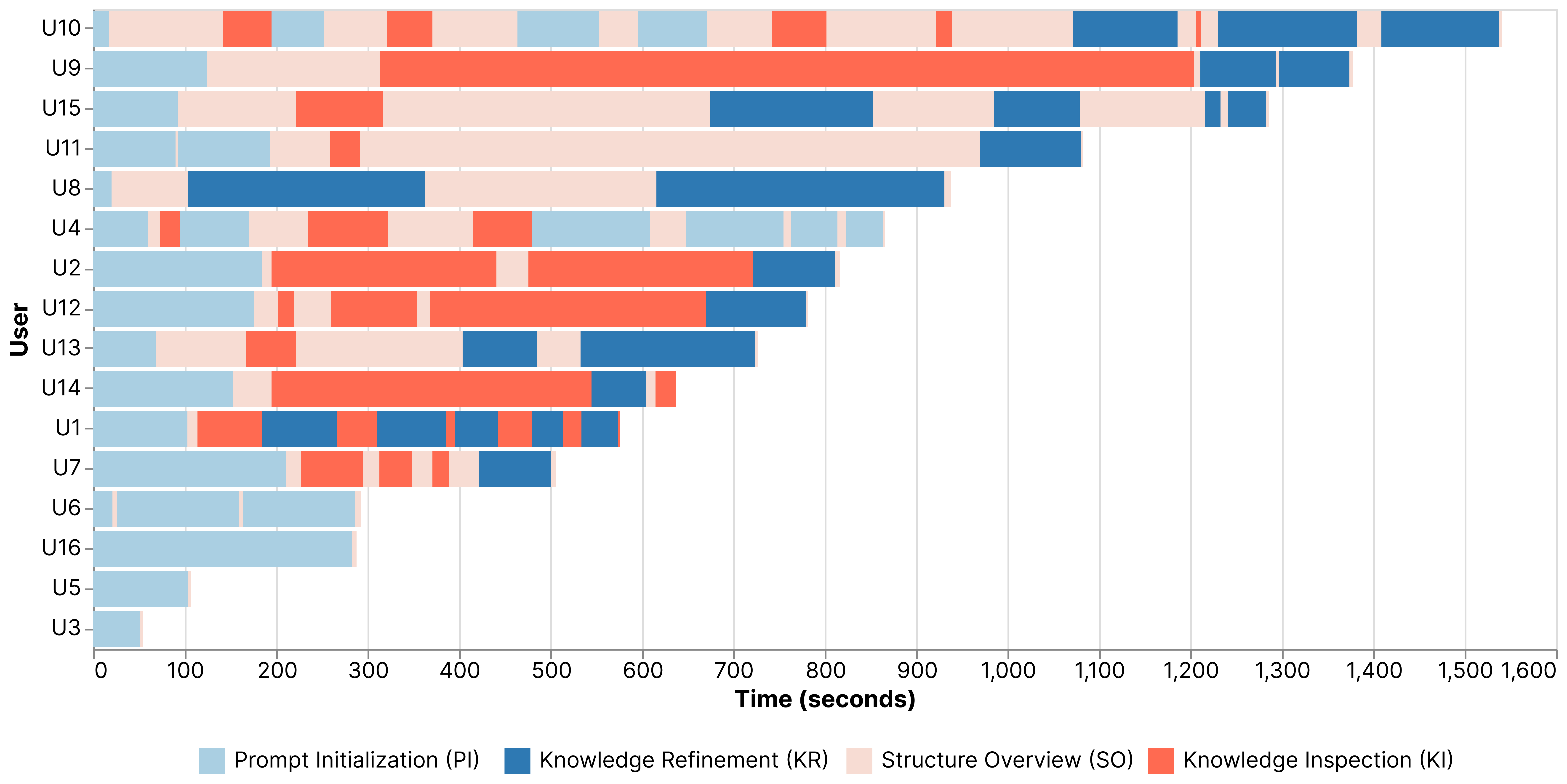}
  \caption{\DIFadd{Interaction timelines for all participants using \mytoolname. Each bar shows one participant's SQL authoring task. Participants may initialize a prompt (\ilogpi{PI}), have an overview of the generated SQL (\ilogso{SO}), inspect the knowledge (\ilogki{KI}), and refine the knowledge (\ilogkr{KR}). Participants are ordered by the task completion time in descending order.}}
  \Description{Interaction timelines for all participants using Cerebra during the SQL authoring task. Stacked horizontal bar chart with time in seconds on the x‑axis (0–1,600) and individual users U1–U16 on the y‑axis, ordered from longest to shortest task completion time. Each user’s bar is segmented into four colored phases: prompt initialization, structure overview of the generated SQL, knowledge inspection, and knowledge refinement. Longer‑duration users (e.g., U10, U9, U15) show extended early segments of prompt initialization and structure overview followed by intermittent blocks of knowledge inspection and refinement. Shorter‑duration users (e.g., U3, U5, U16) show much shorter timelines dominated by prompt initialization with few or no later refinement segments, indicating substantial variability in how extensively participants inspected and refined knowledge before completing the task.}
  \label{fig:cerebra-interaction-logs}
\end{figure*}

\subsection{Results}

\subsubsection{Overall User Experience (RQ1)}

\sloppy

\CAMadd{Most participants (14/16) thought that they could better understand the generated code when using \mytoolname, saying it \usrfb{intuitive} (U4, U5, U7, U8, U12) or \usrfb{clear} (U2, U3, U10, U15).
Many participants (13/16) found that their natural language prompts were shorter and more concise when using \mytoolname.
U2 further remarked, \usrfb{It is more like human communication rather than writing the lengthy code}.
Meanwhile, nearly three-quarters of participants (11/16) said that the code generation of \mytoolname\ was \usrfb{better} compared to Baseline, saying it \usrfb{accurate} (U1-U8, U10) or \usrfb{smart} (U12, U16).
Participants also found it easy to refine the generated SQL.
As U11 said, \usrfb{I don't need to tell it (\mytoolname) which lines of code I want to modify since I can directly select the knowledge that I want to change.}}

\fussy

\CAMadd{These subjective experiences of improved workflow were reflected in objective measures.
Participants completed tasks significantly faster with \mytoolname\ than with Baseline (ANOVA, main effect of tool: $p < 0.01$).
Furthermore, participants reported feeling less burdened when using \mytoolname, rating their perceived Performance ($p < 0.01$) and Effort ($p < 0.05$) workload significantly lower on the NASA-TLX scale compared to Baseline.
Detailed quantitative results are provided in \autoref{appsec:quan-res}.}

\CAMadd{To better understand the actual usage of \mytoolname, we analyzed participants' interaction patterns based on the annotated screen-capture videos.}
\DIFadd{we \CAMdel{first}grouped participants' actions \CAMdel{observed from the screen-capture videos}into two broad categories: \textbf{query construction} and \textbf{code understanding}.
Actions related to \textbf{query construction} include \ilogpi{Prompt Initialization (PI)}, which refers to composing or rewriting the prompt to generate the whole SQL query,
and \ilogkr{Knowledge Refinement (KR)}, which leverages Knowledge View to update the corresponding components of the SQL.
\textbf{Code understanding} actions include \ilogso{Structure Overview (SO)} and \ilogki{Knowledge Inspection (KI)}.
\ilogso{SO} captures interactions with the Script View and its linked Data View to understand the overall structure of the generated SQL,
while \ilogki{KI} corresponds to interactions with the Knowledge View and its linked Data View to check the knowledge.}

\DIFadd{\autoref{fig:cerebra-interaction-logs} reveals several interaction patterns across participants.
First, a small group of participants (U6, U16, U5, U3) completed the task in a very short time and their timelines were dominated by \ilogpi{PI} segments, with almost no time spent on \ilogki{KI} or \ilogkr{KR}.
These participants treated \mytoolname\ primarily as a prompt-based code generator and rarely engaged with the Knowledge View.
Second, several participants (U10, U4, U6) repeatedly returned to \ilogpi{PI} throughout the task, issuing multiple fresh prompts instead of relying mainly on knowledge-level edits, which suggests a preference for resetting the solution when they perceived large mismatches with their intent.
Third, some participants (U9, U15, U11, U2, U12, U14) showed longer continuous periods of \ilogso{SO} and \ilogki{KI} after the first query was generated, followed by shorter \ilogkr{KR} segments.
They invested substantial time in understanding the generated script and its intermediate results before attempting any edits at the knowledge level.
Across these participants, a common pattern was that many of them did not adhere to a single strategy, but instead switched back and forth between \textbf{code understanding} actions and \textbf{query construction} actions as the task progressed.}


\CAMdel{To analyze the user feedback from the interviews, we conducted an inductive content analysis~\mbox{\cite{content-analysis}} and derived the qualitative results as follows:}

\CAMdel{\textbf{Overall user experience}.
Most participants (14/16) thought that they could better understand the generated code when using \mbox{\mytoolname}, saying it ``\textit{intuitive}'' (U4, U5, U7, U8, U12) or ``\textit{clear}'' (U2, U3, U10, U15).}
\DIFdel{For instance, U3 compared the interface of \mbox{\mytoolname} and Baseline by saying, ``\textit{It (\mbox{\mytoolname}) contains more intermediate results in the Knowledge View}''.}
\CAMdel{Many participants (13/16) found that their natural language prompts were shorter and more concise when using \mbox{\mytoolname}.
U2 further remarked, ``\textit{It is more like human communication rather than writing the lengthy code}''.
Meanwhile, nearly three-quarters of participants (11/16) said that the code generation of \mbox{\mytoolname} was ``\textit{better}'' compared to Baseline, saying it ``\textit{accurate}'' (U1-U8, U10) or ``\textit{smart}'' (U12, U16).
Participants also found it easy to refine the generated SQL.
As U11 said, ``\textit{I don't need to tell it (\mbox{\mytoolname}) which lines of code I want to modify since I can directly select the knowledge that I want to change.}''}
\DIFdel{While most participants found \mbox{\mytoolname}'s interface helpful, some participants (3/16) noted that referring to the source code was less convenient.
For example, U9 remarked, ``\textit{The advantage of the second tool (Baseline) is that you can always see the source code in the Chat View, whereas in the first tool (\mbox{\mytoolname}) the code is hidden by default.
If there were an always-on code view, I would feel more trust.}''
U6 added, ``\textit{Combining the strengths of two tools could be better.}''}

\subsubsection{Knowledge-Level Steering and Agency (RQ2)}
\DIFadd{Participants (8/16) described a spectrum of steering strategies across natural language prompts, knowledge refinement, and direct code edits, and shifted between these based on three primary factors: workload, accuracy, and expressiveness.
First, some participants (U5, U9, U10, U15) would roughly assess the workload of different steering strategies, choosing the way that minimized effort given the perceived level of misalignment. 
As U15 explained, \usrfb{If the generated SQL is completely wrong, then I'd definitely just write a new NL instruction and regenerate.
But if it only gets some small parts wrong, then using the knowledge refinement or directly editing the SQL is more convenient.}
Second, historical model accuracy also shaped which channel they used.
After repeated failures of query refinement, participants (U2, U7, U9, U12, U16) became reluctant to keep trying to rephrase the natural language instruction.
Moreover, some participants (U12, U15) reported that certain details of implementation and refinement were hard to describe in natural language, especially when they already had a clear target SQL statement in mind.
As U12 reflected on her experience of repeatedly correcting an incorrect answer through natural language only, 
\usrfb{After I saw the wrong answer, I started thinking about how to phrase my instructions in natural language, so that the system would change its answer into the one I wanted.
Sometimes I felt that process was quite exhausting.}}

\subsubsection{Mental Models of Implicit Knowledge (RQ3)}
\DIFadd{Participants developed mental models of what knowledge the tools possessed by observing the model's behavior and adjusting how explicitly they specified the knowledge in prompts.
Many participants (12/16) began tasks with probing the implicit knowledge boundaries of both tools through trial and error.
As U6 pointed out, \usrfb{I would start with a relatively concise prompt to explore how much it (\mytoolname) already knows.}
When the tool's behavior exceeded their expectations, some participants became more willing to rely on shorter descriptions in subsequent turns.
As U12 commented, \usrfb{I didn't expect it (\mytoolname) to be that smart, but after seeing it perform well, I tended to describe my problem with a much shorter prompt.}
As participants (U1, U4, U5, U12, U13, U16) progressed through the tasks, they formed distinct views about which aspects of their own knowledge had to be spelled out and which could be left implicit, and these views differed across individuals.
For example, as U4 noted, \usrfb{For custom, project-specific concepts, the model (of Baseline) definitely doesn't know them, so I need to explain them;
but for the ones like the concepts of atoms and molecules that are already recognized in a domain, I rely more on the model's own knowledge.}
U5 focused on the contested terminology, \usrfb{If the same norm can have different meanings in different domains, I will clarify it.}
U16 drew a boundary between simple and intricate knowledge:
\usrfb{Things like ratio are often complex, so I can't assume it (Baseline) knows exactly how I want that computed.}}

\subsubsection{Transparency and Trust (RQ4)}
\DIFadd{Several participants (U2, U3, U6, U7, U12) described Baseline as closer to a black box, whereas \mytoolname's Script View and Knowledge View made the structure and intermediate steps more intuitive to inspect.
U3 appreciated the Knowledge View, \usrfb{The tool (\mytoolname) lets me see each small step, with color highlighting that makes it easier to understand.}
Several participants noticed that \mytoolname\ was better at showing the semantics of the query (U1, U6, U7, U12).
As U1 put it, \usrfb{Directly reading code is an inherently inefficient way to check a query.
The strength of the tool (\mytoolname) is not that it makes the code easier to read, but that it more explicitly exposes the execution logic.}
When using \mytoolname, many participants (13/16) relied on the Data View and the metadata attached to knowledge items to validate the correctness of the query, while there was a small number of participants (3/16) still treated raw code as the ultimate source of truth.
For example, U9 remarked, \usrfb{In the first tool (\mytoolname), the code is hidden by default.
If there were an always-on code view like that in Baseline, I would feel more trust.}
Some participants (2/16) further expressed a desire for \mytoolname\ to explicitly display its own knowledge before they entered their prompts, highlighting the need for greater transparency.
As U15 suggested, \usrfb{It (\mytoolname) could use natural language to tell us what knowledge it already has, so that we can, in turn, write our prompts using the terms it is familiar with.}}

\DIFdel{\textbf{Shifts from reviewing code to reviewing knowledge}.
Many participants (12/16) explained why they found code review more effective with \mbox{\mytoolname} by sharing how they leveraged the Knowledge View.
Instead of reading code line by line, participants often started by searching for relevant knowledge items through their titles, which allowed them to quickly narrow down the scope of their review.
For example, as U5 described,
``\textit{If `carcinogenic' is a keyword in my instruction, I will just search the Knowledge View for items whose titles contain words like `carcinogenic', since you cannot directly find the word in the code.}''
After identifying relevant items, participants would read the natural language explanations and metadata to better understand the reasoning behind the model’s outputs and to spot potential issues.
For example, as U16 reported, ``\textit{When I see the metadata `1 record' on an output knowledge item, I immediately realize that the model is wrong}''.
These experiences indicate a shift from low-level code inspection to high-level knowledge review when using \mbox{\mytoolname}.
Participants also provided suggestions for improving the knowledge review process.
As U10 noted, ``\textit{It would be better if the tool (\mbox{\mytoolname}) could intelligently tell me which knowledge item is the most important, so that I can prioritize checking it.}''}

\DIFdel{\textbf{Tendency to reduce explicit knowledge explanation in prompts.}
Through their initial trials, many participants (12/16) observed that even without providing detailed explanations, \mbox{\mytoolname} could still generate relatively accurate results.
Such ``accuracy'' was not necessarily manifested in final execution results, but rather in the execution results in the Knowledge View (e.g., ``\textit{I didn’t expect it (\mbox{\mytoolname}) to be so smart -- out of four knowledge items, it got three right.}'' (U12)).
In such cases, participants preferred to omit explanations of their knowledge from their prompts.
Similarly, U7 mentioned, ``\textit{I look up in the Knowledge View, and find only small fixes are needed, so I prefer to write my initial prompts in a more `vague' way.}''
However, some participants (4/16) still chose to provide detailed explanations of their knowledge in prompts from the start, as they were unsure about the extent of the underlying model's knowledge.
As U13 put it, ``\textit{You can't assume it (\mbox{\mytoolname}) knows everything.}''
Some participants (2/16) further expressed a desire for \mbox{\mytoolname} to explicitly display the knowledge already understood by the underlying model before they entered their prompts, highlighting the need for greater transparency.
For example, as U15 suggested, ``\textit{It (\mbox{\mytoolname}) could use natural language to tell us what knowledge it already knows, so that we can, in turn, write our prompts using the terms it is familiar with.}''}



\section{Technical Evaluation}
\label{sec:tech-eval}
\DIFadd{We conducted a technical evaluation of the performance and robustness of \mytoolname's core components.
While the user study focused on the end-to-end user experience, this evaluation aimed to quantitatively assess the system's model capabilities, specifically the accuracy of knowledge extraction, the precision and recall of knowledge retrieval, and the robustness of the knowledge refinement.}

\subsection{Dataset Preparation}
\label{ssec:dataset-prep}

\DIFadd{Since there are no existing open-source datasets specifically designed for evaluating SQL generation based on historical scripts and implicit knowledge retrieval,
we built a custom dataset derived from BIRD~\cite{bird}.
The custom dataset contains 232 evaluation tasks spanning 4 databases, including \textit{Toxicology} and \textit{European Football}, which were utilized in our user study, as well as two additional databases, namely \textit{Codebase Community} and \textit{Formula 1}.
Each task consists of a natural language description, a ground-truth SQL query, and ground-truth knowledge items to be reused.}

\DIFadd{To balance the data quality and human effort, we employed a two-stage task construction process for each database.
In the first stage, we treated the SQL scripts in BIRD dataset as users' historical queries and randomly sampled 1 to 5 scripts as context.
We used Qwen3~\cite{qwen3}, the same LLM employed in our system implementation (Section~\ref{ssec:sys-impl}), to automatically generate candidate tasks.
These tasks were created under constraints that require the synthesized SQL to be executable, to produce non-empty results, and to maintain a complexity of approximately 30 lines of code, consistent with the difficulty level used in our user study (Section~\ref{ssec:us-apparatus-materials}).
In the second stage, two authors conducted cross‑validation of all tasks, manually verifying the consistency between the natural language description and the SQL query, as well as the correctness of the identified knowledge reuse.
Any issues identified were manually corrected.}

\subsection{Apparatus and Metrics}

\DIFadd{Similar to Section~\ref{ssec:dataset-prep}, we utilized Qwen3 for all procedures involving LLM usage throughout our experiments,
and the all-MiniLM-L6-v2\footnote{\url{https://huggingface.co/sentence-transformers/all-MiniLM-L6-v2}} text embedding model for knowledge retrieval.
Across all system components, we adopted a consistent procedure in which two authors manually cross-validated the semantic correctness of the outputs to determine the metric (e.g., execution accuracy).}

\subsubsection{Knowledge Extraction}
\DIFadd{Direct evaluation of the extraction accuracy is difficult,
because the BIRD dataset provides no ground‑truth knowledge items and the extracted knowledge in \mytoolname's workflow is represented in natural language.
We therefore introduced an indirect evaluation metric, namely \textit{SQL Reconstruction Accuracy}, which measures the execution accuracy of SQL queries reconstructed solely from the extracted knowledge.}

\subsubsection{Knowledge Retrieval and Code Generation}
\DIFadd{For knowledge retrieval, we calculated the precision and recall of retrieved knowledge items.
Here, a knowledge item is defined in the same way as in \mytoolname's workflow (Section~\ref{ssec:implicit-knowledge}) and corresponds to a specific code fragment in the historical SQL scripts.
For code generation, we measured execution accuracy for both direct generation and retrieval-augmented generation.}

\subsubsection{Knowledge Refinement}
\DIFadd{This evaluation used 67 tasks that failed during initial retrieval-augmented code generation.
We define one refinement step as follows: given an incorrect SQL query and its ground-truth counterpart, the model first produces a natural language instruction about code modification, and is then invoked again with only the incorrect SQL and this instruction to revise the original query.
We limited the refinement process to a maximum of 5 steps to avoid non‑convergent cases.}

\subsection{Results}

\subsubsection{Knowledge Extraction Accuracy}
\DIFadd{As shown in \autoref{tab:knowledge-extraction}, the proposed knowledge extraction method demonstrated high accuracy, achieving a success ratio exceeding 95\% across all databases.
The Codebase Community database yielded the highest performance with a 96.77\% success ratio out of 186 total historical queries.
Despite the various domains and size of historical queries, the absolute number of failed tasks remained consistently low at exactly 6 per database.
This consistency indicated that our extraction method could effectively capture the necessary knowledge to construct valid SQL queries for the vast majority of cases.}

\begin{table}[ht]
\centering
\caption{SQL reconstruction accuracy of Knowledge Extraction across four databases.}
\resizebox{0.8\columnwidth}{!}{
    \begin{tabular}{lccc}
    \toprule
    \textbf{Database} & \textbf{History} & \textbf{Success} & \textbf{Success Ratio} \\
    \midrule
    Toxicology & 145 & 139 & 95.86 \\
    European & 129 & 123 & 95.35 \\
    Codebase & 186 & 180 & 96.77 \\
    Formula 1 & 174 & 168 & 96.55 \\
    \bottomrule
    \end{tabular}
}
\label{tab:knowledge-extraction}
\end{table}

\subsubsection{Knowledge Retrieval Performance}
\DIFadd{\autoref{tab:knowledge-retrieval} summarizes the performance of the retrieval module.
The results indicated a design choice for high recall to ensure coverage of relevant knowledge.
Across all databases, the number of retrieved items (``Retrieved'') was consistently higher than the ground-truth knowledge items (``Items''), which naturally diluted Precision (ranging from approximately 43\% to 51\%) but boosted Recall.
The Codebase Community database achieved the highest Recall at 87.67\%, while the European Football exhibited the most balanced performance, achieving the highest Precision (51.63\%) and F1 Score (63.34\%).}

\begin{table}[ht]
    \centering
    \caption{Performance metrics for Knowledge Retrieval.}
    \resizebox{\columnwidth}{!}{
    \begin{tabular}{lccccc}
    \toprule
    \textbf{Database} & \textbf{Items} & \textbf{Retrieved} & \textbf{Precision} & \textbf{Recall} & \textbf{F1} \\
    \midrule
    Toxicology & 334 & 546 & 43.04 & 71.18 & 53.64 \\
    European & 262 & 422 & 51.63 & 81.92 & 63.34 \\
    Codebase & 251 & 514 & 43.49 & 87.67 & 58.14 \\
    Formula 1 & 360 & 645 & 44.74 & 73.20 & 55.53 \\
    \bottomrule
    \end{tabular}
    }
    \label{tab:knowledge-retrieval}
\end{table}

\sloppy

\subsubsection{Impact of Knowledge Retrieval and Refinement}
\DIFadd{\autoref{tab:codegen-refine} presents an ablation study comparing different configurations of the SQL generation pipeline.
The \textit{Direct} generation baseline performed poorly across all databases, with success ratios ranging from 22.22\% to 40.43\%, highlighting the difficulty of generating accurate SQL queries without integration of knowledge.
The introduction of relevant knowledge (\textit{RAG}) yielded a notable improvement, increasing the success rate by 29\% to 47\% across four databases.
Furthermore, the \textit{Pipeline} which combines both RAG and the \textit{Refinement} module, consistently achieved the highest accuracy, surpassing 90\% across all four databases evaluated, with the highest accuracy of 94.03\% on Codebase Community.
The Pipeline success ratio demonstrated the robustness of the Knowledge Refinement module in correcting errors in the generated SQL.}

\fussy

\begin{table}[ht]
\centering
\caption{Ablation study on Code Generation and Knowledge Refinement. \textit{Direct} denotes generation without knowledge retrieval, \textit{RAG} adds retrieval augmentation, \textit{Refine} tests the refinement module independently, and \textit{Pipeline} represents the complete pipeline.}
\resizebox{0.93\columnwidth}{!}{
    \begin{tabular}{llccc}
    \toprule
    \textbf{Database} & \textbf{Mode} & \textbf{Tasks} & \textbf{Success} & \textbf{Success Ratio} \\
    \midrule
    \multirow{4}{*}{Toxicology} 
        & Direct & 55 & 20 & 36.36 \\
        & RAG & 55 & 40 & 72.73 \\
        & Refine & 15 & 10 & 66.67 \\
        & Pipeline & 55 & 50 & 90.91 \\
    \midrule
    \multirow{4}{*}{European} 
        & Direct & 47 & 19 & 40.43 \\
        & RAG & 47 & 33 & 70.21 \\
        & Refine & 14 & 11 & 78.57 \\
        & Pipeline & 47 & 44 & 93.62 \\
    \midrule
    \multirow{4}{*}{Codebase} 
        & Direct & 67 & 27 & 40.30 \\
        & RAG & 67 & 48 & 71.64 \\
        & Refine & 19 & 15 & 78.95 \\
        & Pipeline & 67 & 63 & 94.03 \\
    \midrule
    \multirow{4}{*}{Formula 1} 
        & Direct & 63 & 14 & 22.22 \\
        & RAG & 63 & 44 & 69.84 \\
        & Refine & 19 & 14 & 73.68 \\
        & Pipeline & 63 & 58 & 92.06 \\
    \bottomrule
    \end{tabular}
}
\label{tab:codegen-refine}
\end{table}

\subsubsection{Failure Cases}
\DIFadd{In knowledge extraction, subtle errors may occur even when the correct facts were captured.
Specifically, the extracted descriptions did not always strictly preserve the original output format, which could cause the reconstruction step to change the order of columns or to include unnecessary columns.
Knowledge retrieval was further affected by these imperfections, since in \mytoolname's workflow, the retrieval module matches user natural language instructions with knowledge that is also stored in natural language.
Semantic gaps or differences in phrasing between the two may prevent certain relevant knowledge items from being retrieved, which then led to downstream code generation failures.
Finally, SQL refinement sometimes failed to correct these problems because of hallucinations from the underlying LLM, which may introduce non-existent columns or modify predicates in ways that were not supported by the underlying schema, resulting in refined queries that still executed incorrectly.}

\section{Discussion}
\label{sec:discussion}
In this section, we discuss the implications of our work for NL-to-SQL systems, situate our approach within the context of existing research on knowledge in NL-to-SQL, and outline current limitations and directions for future work.

\subsection{Characterizing and Reusing Knowledge in NL-to-SQL}

\DIFdel{\textbf{Implications}.
In this paper, we propose \mbox{\mytoolname}, an interactive NL-to-SQL tool that aligns implicit knowledge between users and LLMs in two ways.
To make the underlying model capture the knowledge of users, \mbox{\mytoolname} extracted the knowledge in user-written historical scripts based on schemas with column descriptions.
Thus, users can receive more intelligent code suggestions, reducing the burden of repeatedly explaining implicit knowledge when writing prompts.
To make the users understand the implicit knowledge inferred by LLMs, \mbox{\mytoolname} splits the generated code into fragments and offers a Knowledge View to display the knowledge hidden in the code fragments.
This allows users to conveniently check which requirements have been met and which have not.
Such mutual alignment ensures that users and LLMs remain on the same page throughout the coding process, making subsequent code refinements easier.}

\subsubsection{Knowledge for NL-to-SQL}
The notion of ``knowledge'' has received increasing attention in NL-to-SQL research since 2021~\cite{nl2sql-survey}, and has been referred to by different names including external knowledge~\cite{bird, spider2}, domain knowledge~\cite{spider-dk, kaggledbqa}, and ambiguity~\cite{sphinteract, data-ambiguity}.
From the perspective of coverage, ``domain knowledge'' and ``ambiguity'' in prior work mainly include knowledge about dataset-specific conventions, such as abbreviations, conditions, relations between tables, synonyms, and entity-column mappings.
However, they do not sufficiently cover knowledge about task-specific computations, which is also an important part included in implicit knowledge.
From the perspective of knowledge representation, ``external knowledge'' in prior work is typically manifested as unstructured documentations and scripts scattered across the database project, making it challenging to utilize effectively.
Our work categorizes implicit knowledge according to the place it appears in SQL queries (Section~\ref{sssec:types-implicit-knowledge}) to facilitate structured reuse of the knowledge from users' historical scripts.

\DIFdel{\textbf{Generalizability}.
While \mbox{\mytoolname} is designed for aligning implicit knowledge in NL-to-SQL scenarios, our approach have broader applicability to other forms of human-LLM collaboration.
In many creative and analytical domains such as coding, writing, research, or graphic design, users rely heavily on implicit or tacit knowledge that is difficult to verbalize but critical for expert performance.
Recent studies on tacit knowledge in graphic design~\mbox{\cite{tacit-knowledge}} emphasize that much of professional expertise is embedded in intuition, context-specific strategies, and nuanced decision-making, which are not easily codified.
Extending \mbox{\mytoolname}'s knowledge extraction and alignment mechanisms to these domains can enable LLM-driven systems to better capture tacit knowledge from user actions, historical artifacts, and non-verbal cues.
This could facilitate richer knowledge alignment, reduce repeated clarification, and foster more effective human-AI co-creation. 
Furthermore, by making tacit knowledge explicit and actionable through structured representations and action tracking, future systems can support not just code generation but also collaborative design, writing, or research workflows, paving the way for generalizable frameworks of knowledge-driven human-LLM interaction.}

\subsubsection{Reuse as a Mechanism for Semantic Alignment in Data Analysis}
\DIFadd{Reuse has long been established as a fundamental strategy in data analysis tools to reduce authoring effort and lower technical barriers.
Prior work has successfully demonstrated the value of reuse in generating infographics~\cite{retrieve-then-adapt, metaglyph}, authoring data analysis code~\cite{edassistant}, recommending visualizations~\cite{solas}, and constructing data reports~\cite{respark}.
These approaches typically accelerate creation by reusing explicit artifacts, such as visual templates, code snippets, or narrative structures.
\mytoolname\ extends this paradigm to the domain of database querying, but shifts the emphasis of reuse from the structure or style to semantics.
By treating historical scripts as a repository of codified conventions, \mytoolname\ leverages reuse to ground the LLM's generation in the user's specific context.
This shift suggests that in AI-assisted data analysis, the value of reuse lies in stabilizing the stochastic nature of LLMs by anchoring them to the verified logic embedded in the user's interaction history.}

\subsection{Knowledge Management in NL-to-SQL Generation}

\subsubsection{Deciding What Types of Knowledge to Make Explicit, Traceable, or Ephemeral}
\DIFadd{In high-stakes contexts such as finance or manufacturing, misapplied knowledge can lead to financial losses or legal liability.
In these settings, knowledge on computational logic needs to be made fully explicit and fixed, rather than inferred by embeddings or LLMs.
A future version of \mytoolname\ could include alert mechanisms that flag when unverified knowledge is being used in such sensitive computations.
A second category involves knowledge that must be traceable, for example norms and formulas whose authorship and temporal validity matter in multi-stakeholder environments.
Here, who introduced a convention and when it was last updated are as important as the rule itself, such as interest rate parameters that change over time.
In the future, \mytoolname\ could be extended with auditing features that associate knowledge items with provenance and version histories.
Finally, for the tasks with low risk, it is often more efficient to let LLMs generate task-specific computations ephemerally without the overhead of formalization, as long as users can revise the resulting knowledge when needed.}

\sloppy

\subsubsection{Maintaining Knowledge Validity under Schema Evolution}
\DIFadd{Since \mytoolname\ extracts knowledge from specific historical code snippets, changes in table structures, column names, or data values may cause previously extracted knowledge to become partially stale or even misleading.
One potential approach to improving the resilience of \mytoolname\ is to establish links between the entities in these snippets (e.g., table and column names) and the live database schema.
When schema updates occur, the system can identify the knowledge items impacted by the modified entities and automatically regenerate their corresponding SQL snippets as well as natural language descriptions.
Future work can further investigate such incremental update mechanisms and evaluate their effectiveness in preserving the reliability of extracted knowledge.}

\fussy

\subsection{Balancing LLM Automation and Human Intervention}

\DIFadd{To balance cognitive load against query precision, \mytoolname\ hands off labor-intensive information processing tasks to the LLM.
Such tasks involve column description generation in the data dictionary and knowledge extraction from massive SQL scripts, where manual processing is prohibitively expensive.
Conversely, tasks like final SQL generation that directly determine the correctness of data queries, are designated to be verified and crafted by humans through the interface.
However, errors in the upstream LLM automation (e.g., inaccurate column descriptions or extracted knowledge) can cascade~\cite{chainforge}, degrading the accuracy of downstream retrieval and generation.
Yet, requiring users to manually inspect every generated description or extracted knowledge item is impractical, especially for ad-hoc queries where efficiency is paramount.}

\DIFadd{To address this trade-off between automation efficiency and error propagation, a compromise is to employ indirect metrics to rapidly localize LLM errors without exhaustive manual review~\cite{chainforge, evallm, who-validates-the-validators}.
For instance, our technical evaluation assesses the quality of knowledge extraction by measuring the accuracy of reconstructing SQL queries from the extracted knowledge, using reconstruction failure as a proxy for extraction error.
In the future, we aim to generalize this approach by adopting frameworks like EvalGen~\cite{who-validates-the-validators} and EvalLM~\cite{evallm}, allowing users to define high-level criteria or metrics (e.g., ``\textit{consistency with naming conventions}'') to automatically audit the LLM's outputs.
By shifting the human role from reviewing individual outputs to interactively defining and validating these evaluation criteria, \mytoolname\ can better foster user trust and interpretability for such labor-intensive information processing while minimizing the manual effort required for verification.}

\subsection{Threats to Validity}

\DIFadd{In this section, we discuss the threats to validity for our preliminary study and user study.}

\DIFadd{For the preliminary study, we acknowledge that participants' individual workflows and prior experience with NL-to-SQL tools could introduce bias, potentially influencing the interview results.
To mitigate this threat, we recruited a cohort of 10 data practitioners from diverse industrial domains, ensuring that our findings were not overfitted to a specific data schema or workflow.}

\DIFadd{For the user study,
we recognize that participants performed tasks on domain-specific datasets (Sports, Chemistry) with which they lacked prior familiarity.
However, this design aligns with our goal to evaluate \mytoolname\ for a broad spectrum of data practitioners, regardless of their specific domain background. 
We argue that implicit domain knowledge consists of two distinct facets:
task-specific computations (e.g., scientific formulas) and dataset conventions (e.g., how attributes are stored and linked).
While end-user scientists may possess the former, general data practitioners frequently face the challenge of navigating dataset conventions to write accurate SQL.
For instance, calculating a ``student's average grade'' requires not just the arithmetic logic but also understanding the specific schema conventions for locating and aggregating those grades.}

\DIFadd{Therefore, our study tasks were designed to encompass both conventions and computations.
To bridge the knowledge gap, we provided tutorial training that enabled participants to grasp the necessary context.
The results showed that no participant failed due to a lack of domain knowledge, validating this approach.
However, we acknowledge that evaluating \mytoolname\ with domain experts like end-user chemical scientists could provide a deeper understanding of how implicit domain conventions are externalized and reused during SQL authoring.
Such an evaluation poses additional challenges such as recruiting sufficient domain experts, designing comparable tasks across conditions, and disentangling the effects of SQL skill from domain familiarity.
We leave such a study to future work to further extend the external validity of our findings.}

\subsection{Limitations and Future Work}

We observe two limitations regarding the functionalities of \mytoolname.
\DIFdel{First, when users set up the knowledge extraction process, the data dictionary interface (\mbox{\autoref{fig:data-dict}}) still requires users to manually check each column description.
When the number of columns in a database increases, this can impose a considerable burden on users.
To further reduce users' effort in reviewing column descriptions, future work could incorporate a confidence estimation mechanism, allowing users to prioritize checking the column descriptions whose meanings the LLM is less certain about.
Second,}\DIFadd{First,} the knowledge extracted from historical scripts in \mytoolname\ is tailored to a single database schema and does not generalize well to other schemas.
To address this, future work could explore transferring knowledge between datasets with semantically similar schemas.
\DIFadd{Second, \mytoolname\ is currently implemented as a standalone web application, which does not fully align with SQL developers' workflows in desktop IDEs.
This separation may introduce friction and limit \mytoolname's access to SQL artifacts typically managed within IDE environments and version control systems.
Future work could integrate \mytoolname\ as an IDE extension, potentially embedding the Script View within the editor panel while exposing the Knowledge and Data Views as lightweight, expandable side panels.
This would provide a more seamless user experience and enable tighter coupling with developers' historical SQL contexts.}

\section{Conclusion}
\label{sec:conclusion}
Data practitioners often provide natural language instructions involving implicit knowledge and struggle to validate whether NL-to-SQL tools have correctly inferred their requirements.
In this paper, we present \mytoolname, an interactive NL-to-SQL tool that aligns implicit knowledge between users and LLMs throughout SQL script generation and refinement.
We began with a preliminary study to analyze the practices and challenges users encounter with existing NL-to-SQL tools, summarizing three key user requirements through interviews.
Building on these findings, \mytoolname\ automatically extracts and reuses implicit knowledge from users’ historical scripts, provides a knowledge tree view to make the implicit knowledge explicit, and supports knowledge-level iterative query refinement.
Our user study with 16 data practitioners demonstrates that \mytoolname\ improves efficiency and helps users review the code more effectively.
The notion of implicit knowledge and the knowledge reuse techniques can be generalized to other human-AI collaboration tasks.
In the future, we plan to explore automatic confidence estimation and investigate cross-database knowledge transfer to further enhance the usability and generalizability of \mytoolname.

\begin{acks}
The work was supported by Zhejiang Provincial Natural Science Foundation of China under Grant No. LD25F020003, NSFC (62402421, 62421003), and Ningbo Yongjiang Talent Programme (2024A-399-G).
Dae Hyun Kim is partially supported by the grant 2025-22-0499 awarded by the Institute for AI and Social Innovation at Yonsei University.
We sincerely appreciate the constructive feedback from the anonymous reviewers and all participants in our studies.
\end{acks}

\bibliographystyle{ACM-Reference-Format}
\bibliography{main}

\clearpage
\appendix


\onecolumn

\section{Demographic Information of participants}

\begin{table*}[htbp]
\centering
\caption{Demographic Information of Participants in Preliminary Study.}
\label{tab:demographics-preliminary}
\begin{tabular}{ccccccc}
\hline
User & Gender & Age & Background & \makecell{Experience\\(Year)} & \makecell{Familiarity\\(1--7)} & Frequency \\
\hline
P1 & Male & 26 & Enterprise Resource Planning & 7 & 7 & Once a week \\
P2 & Male & 41 & Natural Resources & 10 & 2 & Everyday \\
P3 & Male & 41 & Geographic Information System & 20 & 6 & Once a week \\
P4 & Male & 28 & Stock Market Analysis & 3 & 3 & Everyday \\
P5 & Female & 34 & Education & 10 & 4 & Three times a week \\
P6 & Male & 32 & Natural Resources & 9 & 5 & Everyday \\
P7 & Male & 39 & Geospatial Data Query & 20 & 7 & Everyday \\
P8 & Male & 40 & Land Affairs & 17 & 3 & Everyday \\
P9 & Male & 37 & Geographic Information System & 5 & 4 & Once a week \\
P10 & Male & 38 & Communications & 14 & 6 & Everyday \\
\hline
\end{tabular}
\end{table*}

\begin{table*}[htbp]
\centering
\caption{Demographic Information of Participants in User Study.}
\label{tab:demographics-user}
\begin{tabular}{ccccccc}
\hline
User & Gender & Age & Background & \makecell{Experience\\(Year)} & \makecell{Familiarity\\(1--7)} & Frequency \\
\hline
U1 & Female & 26 & Cloud Computing & 3 & 7 & Everyday \\
U2 & Male & 21 & E-commerce Management Systems & 5 & 7 & Once a week \\
U3 & Female & 22 & Gaming, Biology & 1 & 5 & Once a week \\
U4 & Male & 27 & Stock Database Query & 5 & 7 & Once a week \\
U5 & Male & 24 & Chemical Engineering & 1 & 3 & Once a week \\
U6 & Male & 37 & Surveying & 5 & 4 & Once a week \\
U7 & Male & 26 & Knowledge Graph Query & 5 & 7 & Once a week \\
U8 & Female & 20 & Pharmaceutical Retail & 1 & 5 & Three times a week \\
U9 & Male & 40 & Office Automation Data & 17 & 3 & Everyday \\
U10 & Male & 26 & E-commerce & 7 & 7 & Once a week \\
U11 & Male & 20 & Geoinformation Science & 1 & 4 & Once a week \\
U12 & Female & 23 & Banking & 1 & 3 & Three times a week \\
U13 & Female & 26 & Internet Backend Development & 5 & 5 & Everyday \\
U14 & Male & 23 & Bike Sharing & 2 & 5 & Everyday \\
U15 & Male & 28 & Financial Data Verification & 3 & 3 & Everyday \\
U16 & Female & 23 & \makecell{Application Commercialization} & 1 & 2 & Three times a week \\
\hline
\end{tabular}
\end{table*}

\clearpage
\twocolumn

\section{Quantitative Results of User Study}
\label{appsec:quan-res}

\begin{figure}[th]
  \centering
  \includegraphics[width=\columnwidth]{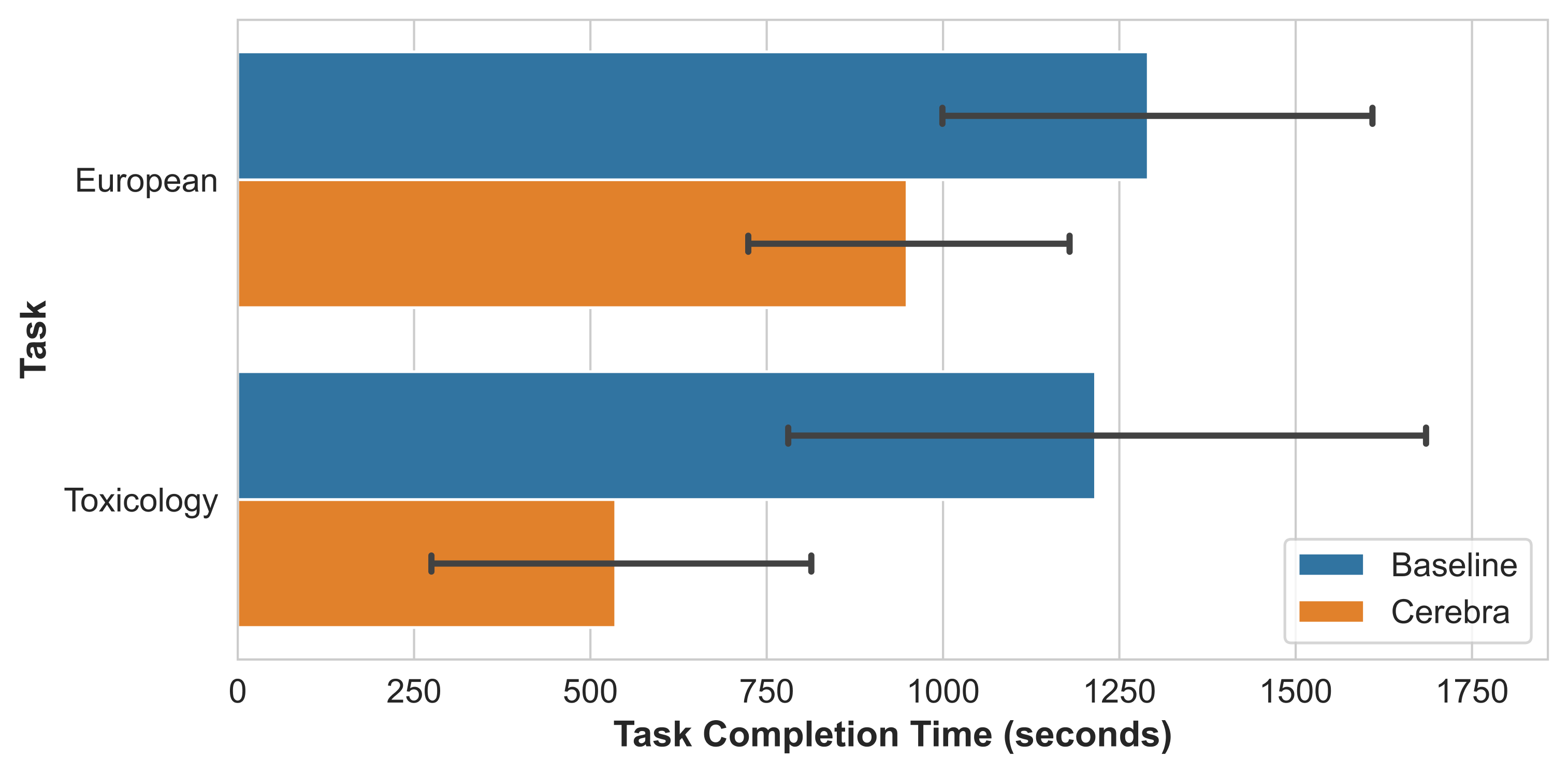}
  \caption{\DIFadd{Task completion time for Baseline and \mytoolname.}}
  \Description{A horizontal bar chart showing task completion times in seconds for the “European Football” and “Toxicology” tasks, where Cerebra results in notably shorter completion times than Baseline for both tasks, with smaller error bars for Cerebra}
  \label{fig:comptime}
\end{figure}

\CAMadd{To evaluate whether \mytoolname\ accelerated the script authoring process, we measured ``task completion time'' of each participant in each task.
Shapiro-Wilk and Levene’s tests confirmed that the raw data followed a normal distribution with equal variance.
Consequently, we performed a two-way ANOVA to test the main effects of tools (\mytoolname\ vs. Baseline) and tasks (European vs. Toxicology), as well as their interaction effects.
We found that the main effects of tools were significant ($p < 0.01$), while the interaction effects and the main effects of tasks were not significant.
According to the result of task completion time (\autoref{fig:comptime}), participants spent less time completing the tasks with \mytoolname\ than with Baseline, reflecting the overall advantage indicated by the ANOVA.}

\begin{figure}[th]
  \centering
  \includegraphics[width=\columnwidth]{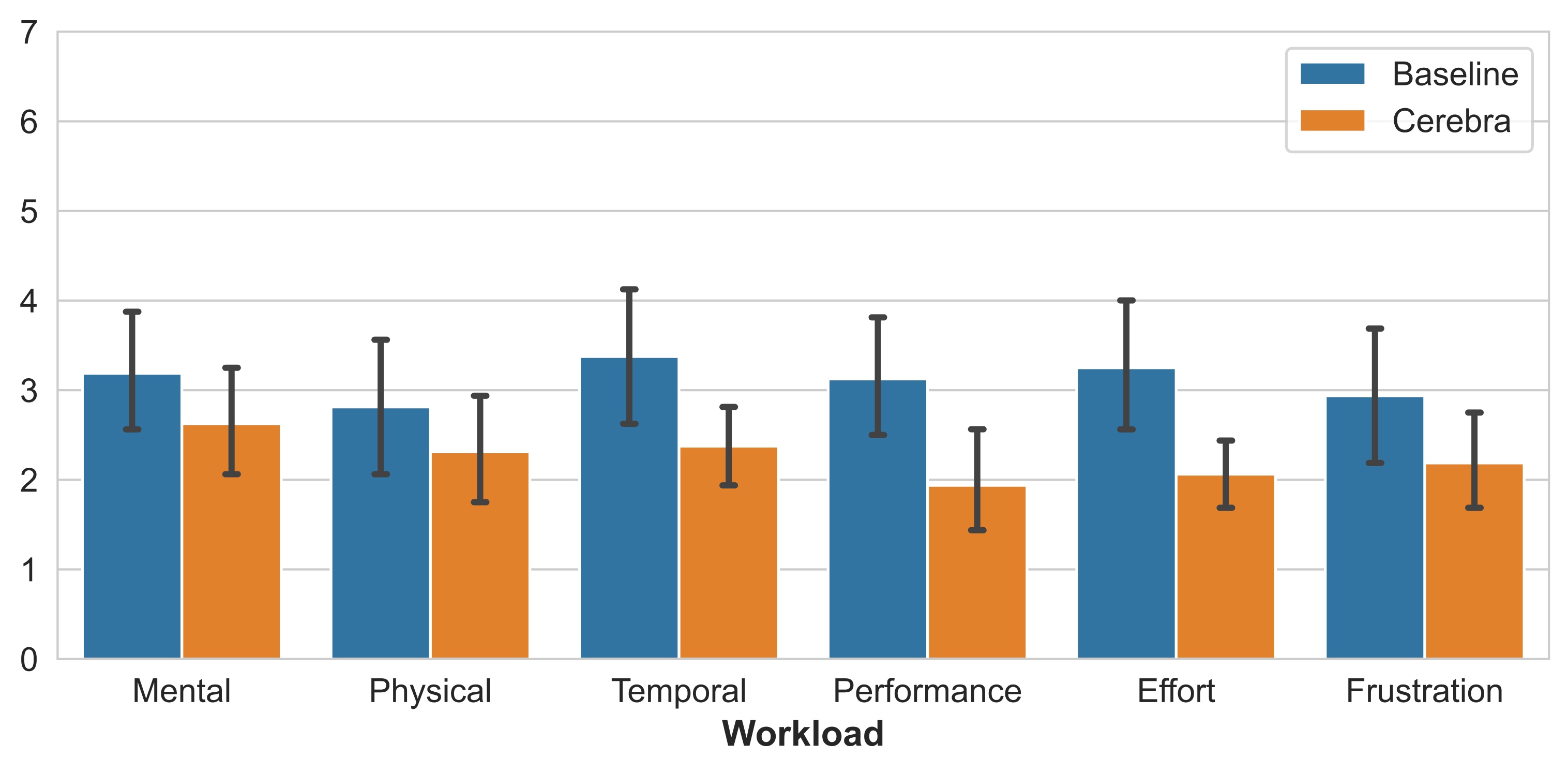}
  \caption{\DIFadd{NASA‑TLX workload for Baseline and \mytoolname.}}
  \Description{A grouped bar chart comparing subjective workload ratings on six dimensions—Mental, Physical, Temporal, Performance, Effort, and Frustration—where Cerebra has lower average ratings than Baseline across all categories, indicating reduced perceived workload, though the error bars show some overlap between conditions.}
  \label{fig:workload}
\end{figure}

\CAMadd{For perceived workload, we analyzed the NASA‑TLX data, as shown in \autoref{fig:workload}.
Normality tests indicated that only the Mental Demand dimension met the assumptions for a standard ANOVA.
For the remaining dimensions (Physical Demand, Temporal Demand, Performance, Effort, and Frustration), we utilized the non-parametric Aligned Rank Transform (ART) ANOVA. 
The analysis revealed significant main effects of tools on the Performance ($p < 0.01$) and Effort ($p < 0.05$) workload scores.
Post‑hoc EMMeans comparisons confirmed that participants reported significantly lower Performance and Effort workload when using \mytoolname\ than when using Baseline.
All other NASA‑TLX dimensions (Mental, Physical, Temporal, and Frustration) showed no significant effect of tools.
Similar to the time analysis, neither the main effect of tasks nor the interaction between tools and tasks was significant across any dimension.}

\end{document}